# Data-based *Polymer-Unit Fingerprint (PUFp)*: A Newly Accessible Expression of Polymer Organic Semiconductors for Machine Learning

***Xinyue Zhang [1], Genwang Wei [1, 2], Ye Sheng [3], Jiong Yang [3, 4], Caichao Ye [1, 2, *], Wenqing Zhang [1, 2, *]***

[1] *Department of Physics & Guangdong Provincial Key Laboratory of Computational Science and Material Design, Southern University of Science and Technology, Shenzhen, 518055, PR China*

[2] *Academy for Advanced Interdisciplinary Studies & Department of Materials Science and Engineering, Southern University of Science and Technology, Shenzhen, 518055, PR China*

[3] *Materials Genome Institute, Shanghai University, Shanghai, 200444, PR China*

[4] *Zhejiang Laboratory, Hangzhou, 311100, PR China*

[*] *Corresponding Authors: C. Ye: yecc@sustech.edu.cn, W. Zhang: zhangwq@sustech.edu.cn.*

**Abstract**

In the process of finding high-performance organic semiconductors (OSCs), it is of paramount importance in material development to identify important functional units that play key roles in material performance and subsequently establish substructure-property relationships. Herein, we describe a *polymer-unit fingerprint (PUFp)* generation framework. Machine learning (ML) models can be used to determine structure-mobility relationships by using *PUFp* information as structural input with 678 pieces of collected OSC data. A polymer-unit library consisting of 445 units is constructed, and the key polymer units for the mobility of OSCs are identified. By investigating the combinations of polymer units with mobility performance, a scheme for designing polymer OSC materials by combining ML approaches and *PUFp* information is proposed to not only passively predict OSC mobility but also actively provide structural guidance for new high-mobility OSC material design. The proposed scheme demonstrates the ability to screen new materials through pre-evaluation and classification ML steps and is an alternative methodology for applying ML in new high-mobility OSC discovery.

## Table of Contents

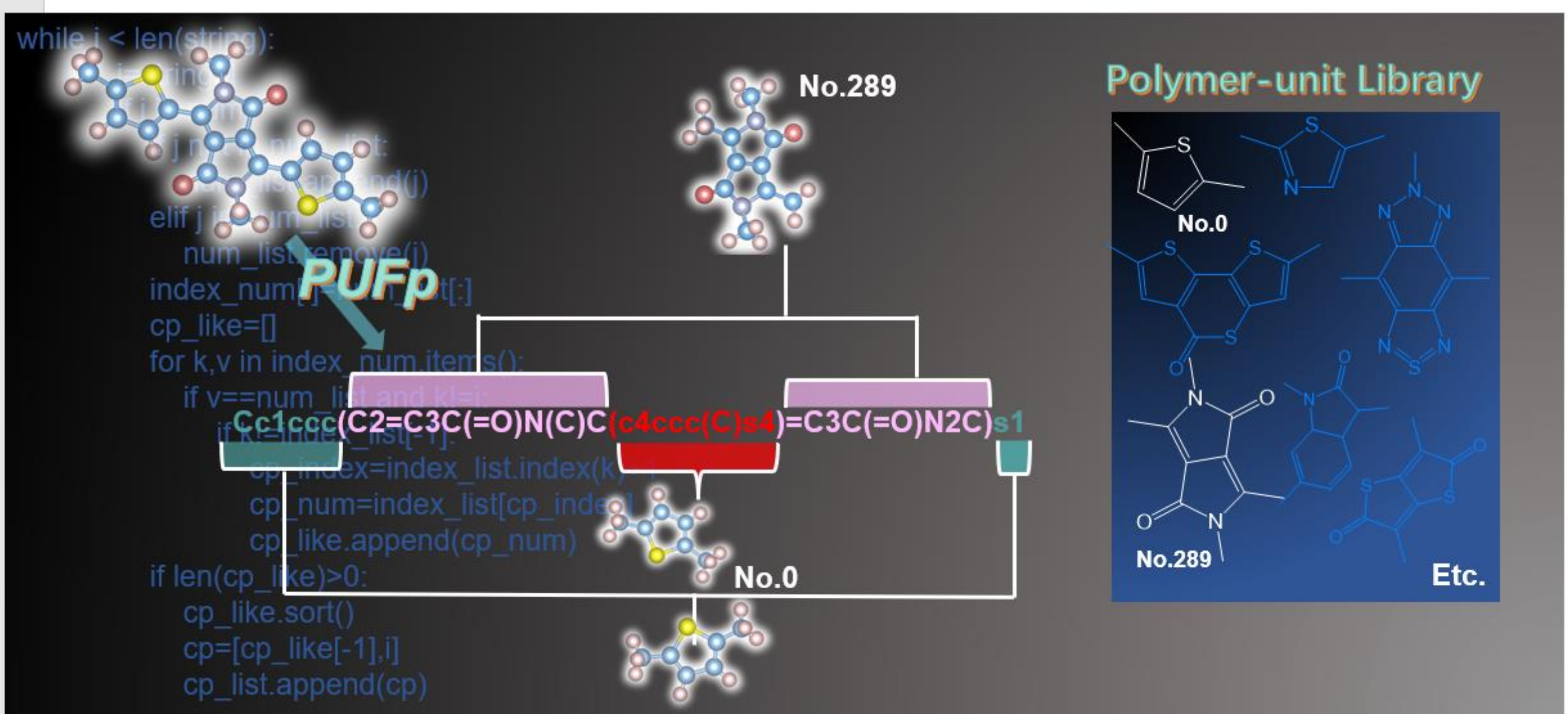

## 1. Introduction

Organic semiconductors (OSCs) are widely applied in a variety of areas, such as photovoltaic materials [1-2], field-effect transistors [3-6] and thermoelectric materials, due to their characteristics of low synthesis costs, light weight structures, and strong flexibility and plasticity attributes [7]. Because many properties of OSCs are related to carrier transport, the design and synthesis of novel materials with high carrier mobility has become a key factor in the development of high-performance optical and electrical devices. Semiconductors can be divided into n-type, p-type and bipolar materials; the carriers of p-type semiconductors are holes, and those of n-type semiconductors are electrons [8]. Recent studies have shown that the mobility of polymer semiconductors can be improved by adjusting branched chains [9-10], combining different organic functional groups (D-A, D-D, A-A. etc.)[11] or doping with small molecules [12]. The carrier mobility of p-type semiconductor materials has been reported to be as high as 71 $cm^2 \cdot V^{-1} \cdot s^{-1}$ [10], and the mobility of n-type semiconductors can also reach 14.4 $cm^2 \cdot V^{-1} \cdot s^{-1}$ [13]. In recent years, the mobility records of OSCs have continued to advance [14-19], which proves that there is plenty of room for improvement in the mobility of OSCs.

Machine learning (ML) can train advanced algorithms based on big data to capture intrinsic connections that may embody underlying principles; thus, ML algorithms have led to significant progress in many fields [20-22]. In materials research and development, including studies on catalyses [23-25], batteries [26-29] and OSCs [30-32], ML is able to construct structure-property material atlases and to design new materials [33], especially in fields where properties and structures are closely related. When ML is used for the construction of structure-property material atlases, the way in which structural information of materials is expressed determines which material descriptors or features can be studied. Recently, scientists have commonly used molecular fingerprints [34], the most convenient digital descriptor converted from a string of certain chemical structures (simplified molecular input line entry system: SMILES) using a tool equipped with a computer language platform, to express the structure of organic molecules for ML [35]. In 2019, Sun et al. [36] explored several ways to express molecular structures, i.e., images, ASCII strings, descriptors, and fingerprints, as inputs for various ML algorithms and studied their effect on the prediction of power conversion efficiency (PCE), which is a key performance indicator of optoelectronic materials. It was found that a *hybridization* fingerprint with a length of over 1000 bits combined with a classification model trained by the random forest (RF) algorithm can obtain high prediction accuracy (up to 81.76% N/P classify - N-type: electronic mobility>hold mobility; P-type: electronic mobility < hold mobility). This high accuracy indicates that molecular fingerprints, as inputs for the expression of molecular structures used in ML, have good predictive performance for organic polymer materials.

Molecular fingerprints, such as *FP2* [37], *Morgan* [38], *and Torsion* [39], have been designed for large-scale database screening and take the form of an array of bits. They contain "1"s and "0"s to describe the presence or absence of particular substructures/patterns in a molecule and can be used as inputs to train ML models to predict the properties of molecular materials. *Morgan* fingerprints are circular fingerprints that are generated by the Morgan algorithm and represent substructures within a certain distance from the target atom [38] **(Figure 1a)**. The *Torsion* fingerprint **(Figure 1b)** is a path-based fingerprint, where each of the 32 digits represent a linear path of 4 nonhydrogen atoms, and contains π electron numbers, atom types, and nonhydrogen branch chain numbers [39]. The *FP2* molecular fingerprint is also a path-based molecular fingerprint that indexes small molecular fragments based on a linear path of seven atoms [37]. However, the direct interpretation of these ML models that use these fingerprints as input is difficult, and the specific substructures that greatly influence the performance cannot be well identified. If each bit of a molecular fingerprint represents a specific substructure, important substructures that greatly influence the target performance can be identified by analysis and reverse-trained ML models; for example, *MACCS* [40] and *La FREMD* [41] fingerprints implemented by *RDKIT* [42] can be used to identify key substructures, applied in organic photovoltaics (OPV) material design, and screened by ML **(Figure 1 d)**. However, the disadvantage of the *MACCS* fingerprint **(Figure 1 c)** is that each bit represents a very small chemical structure (such as a double bond, a hydroxyl group, a carbonyl group, an *S−N* group, or an alkaline metal), and for the *La FREMD* fingerprint, each bit represents a randomly cropped molecular fragment according to certain rules. Obviously, if substructures are not a common functional structure, such as polymer units in OSCs, it is difficult to apply the identified key substructures to construct new materials and explore the relationship between structure and properties.

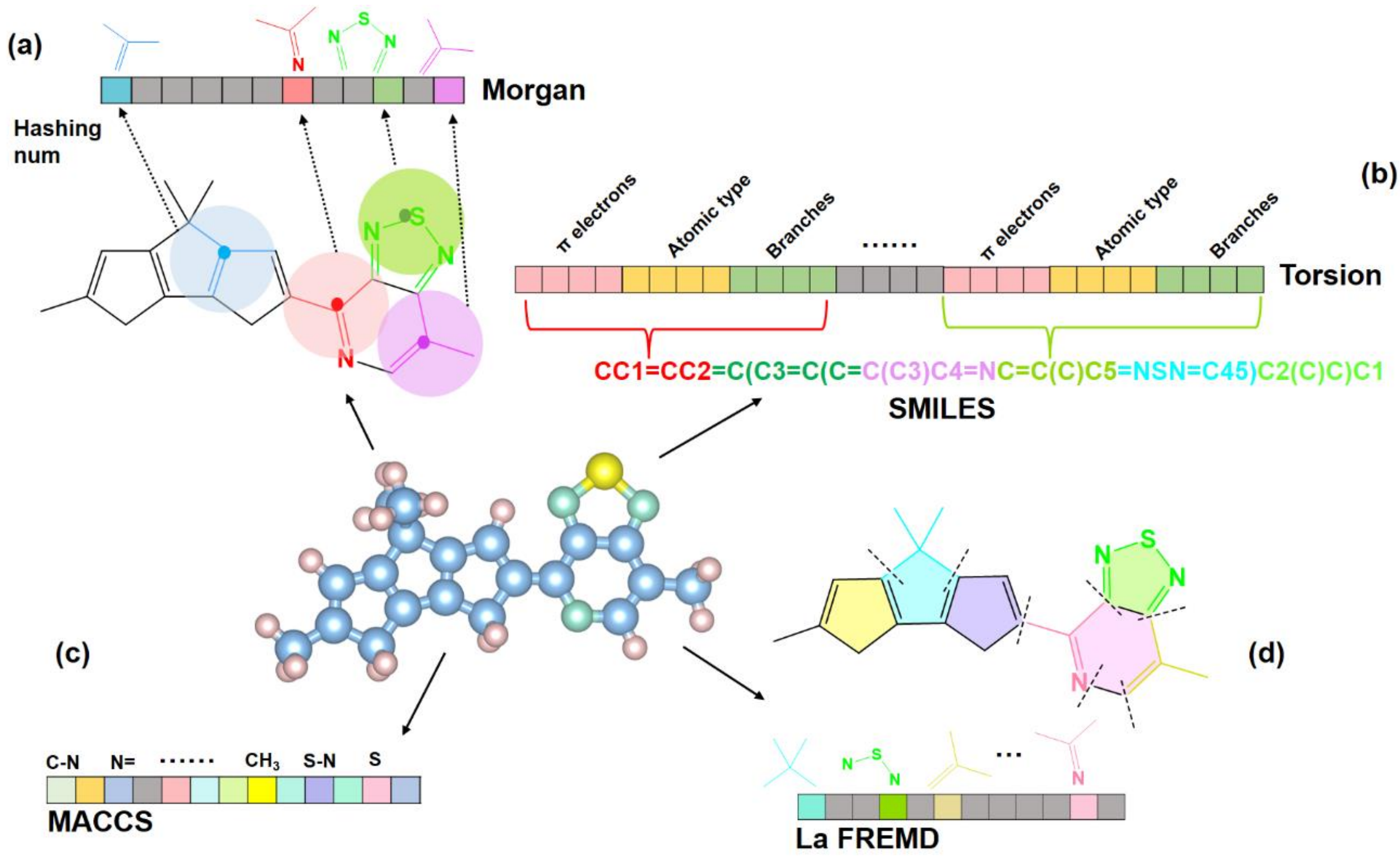

**Figure 1.** Schematic diagrams of several molecular fingerprints. (a) ***Morgan***: *Morgan* fingerprints consist of hash indexes of substructures within a certain distance of each atom in the structure. (b) ***Torsion***: *Torsion* fingerprints are linear paths of 4 atoms at a time, and the number of π electrons, atomic species, and nonhydrogen branch chains of these 4 atoms are recorded in the fingerprints. (c) ***MACCS***: Each bit in the *MACCS* code represents a small substructure and adopts one-hot encoding. (d) ***La FREMD***: The *La FREMD* fingerprint generated by *RDKIT* is a molecular fragment generated by clipping the input according to certain rules. One-hot coding was used to represent each molecular fragment.

In this work, a data-based dynamic *polymer-unit fingerprint (PUFp)* was designed and proposed to segment suitable functional structure units to represent each bit of a molecular fingerprint. All organic monomers are bonded by polymer units (mono rings, bicyclic rings, fused rings and branched chains). Therefore, although organic monomers are multitudinous, polymer units are common and restricted. The *PUFp* generation framework is schematically illustrated in **Figure 2**. The organic polymers and units are dynamically generated according to a user-provided database through the Python-based *polymer-unit-recognition script* (*PURS,* access by this link: https://github.com/xinyue123-q/Python-based-polymer-unit-recognition-script-PURS-) or see the **Supporting Information**), and each bit represents a polymer unit. We demonstrate that *PUFp* can be utilized in the entire process during the investigation of polymer OSCs by ML. The *N/P* classification accuracy of an OSC is above 80%. Through the *Shapley Additive exPlanations* (*SHAP*) analysis [43-44] of trained ML models, the key polymer units that greatly influence the mobility of OSCs could be identified. A polymer-unit library that consists of 445 units is created, and the top-positive-important units for *n*- or *P*-type mobility and *N/P* classification are discovered. The relationship and mechanism between the OSC polymer units and their combination modes with experimental carrier mobilities are also investigated and studied. Through this work, the application mode of fingerprints changes from "black" to "white", i.e., switching from a passive "black box" model construction to active mechanism exploration and a new material design. As a result, the ML method combined with *PUFp* can fundamentally change the methodology for new organic material discovery.

## 2. Methods

### 2.1 Summary of the polymer-unit fingerprint (*PUFp*) generation framework for developing high-mobility OSC materials with ML.

The *PUFp* generation framework and the workflow for its application in developing high-mobility OSC materials with ML are schematically illustrated in **Figure 2**. It consists of six integral components that perform four tasks separately:

I. Starting with the establishment of a polymer OSC materials dataset, which consists of 678 different polymer OSC materials that have been experimentally reported (**Figure 2a** and SI 1. Polymer-unit SMILES). The N/P type distribution of these data is shown in **Figure S1**. We analysed the SMILES of each OSC and divided each polymer unit by using our *PURS* framework (**Figure 2b**), and the polymer-unit library was created (**Figure 2c**). Here, the polymer units are updated according to the polymer OSC database, so the polymer unit library is updated dynamically based on the OSC database.

II. Each polymer OSC SMILES in the polymer OSC dataset (**Figure 2a**) was indexed by the polymer-unit library (**Figure 2c**) to generate *PUFp* in one-hot form (**Figure 2d**). First, a node matrix is generated according to the polymer-unit index in the database and the polymer-unit library, and each node can contain one polymer unit. The number of columns in the node matrix represents the maximum number of nodes $N$ according to the OSC database, and the number of rows represents the number of polymer-unit types $T$ according to the polymer-unit library. Then, we sum each row of the node matrix and obtain a one-dimensional vector, which we named *PUFp*.

III. Five commonly used ML algorithms are employed by the self-built polymer OSC materials database and our newly designed *PUFp* framework, which can express 445 different polymer units of an organic molecule, to implement property prediction model construction and unit importance analysis. With this targeted feature engineering, we can identify the key basic polymer units that greatly influence the OSC mobility performance (**Figure 2e**).

IV. A new material virtual library is generated by rearranging the most important polymer units based on the three main types of polymer units (mono rings, bicyclic rings, fused rings and branched chains; see **Figure 2f**). The ML model trained in the previous step is used to predict the mobility performance of the new material virtual library, and the new materials are classified into high, medium, and low types. The polymer-unit type combinations with the highest proportion of the high-mobility types are calculated, and the favoured polymer-unit combinations for highly mobile OSCs are found.

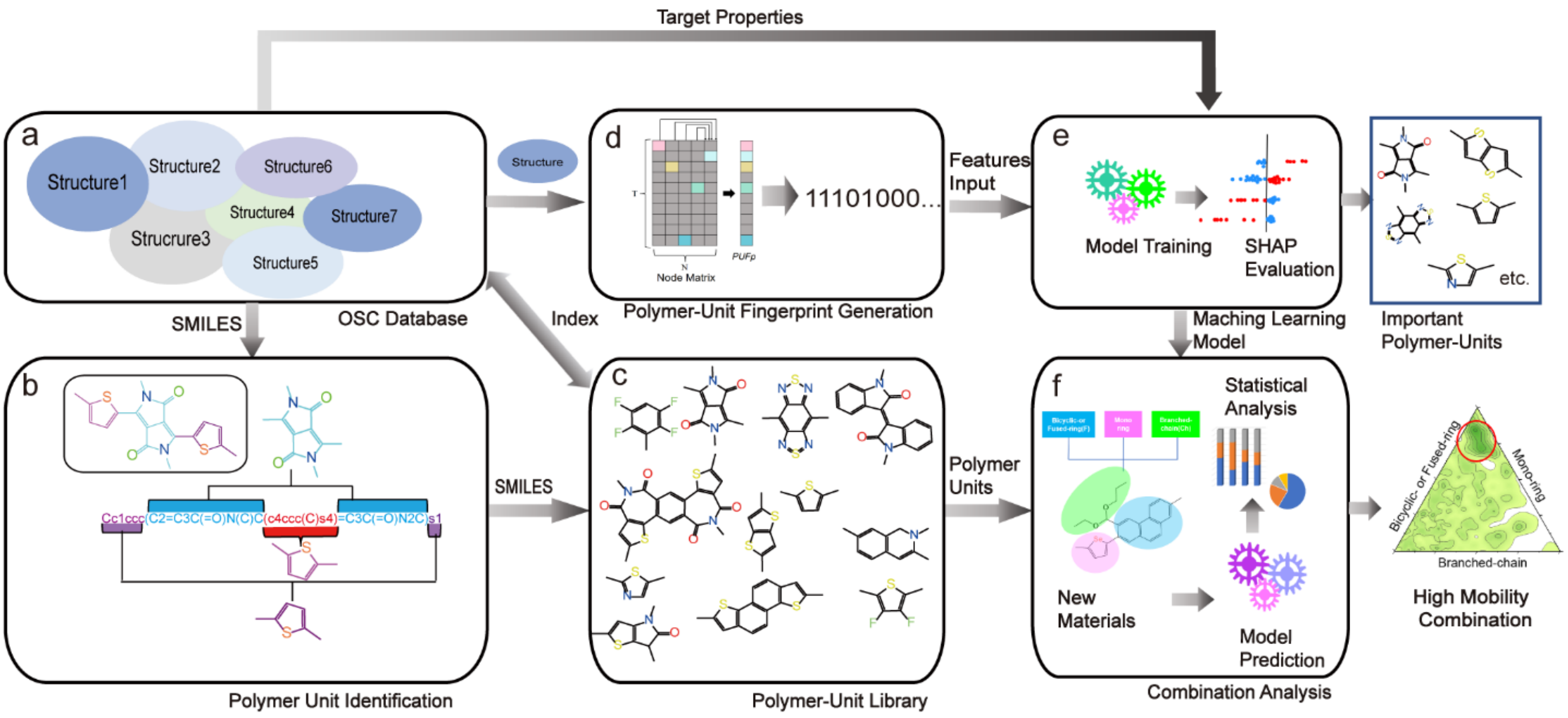

**Figure 2**. Scheme of the *PUFp* generation framework and the workflow for its application in developing high-mobility OSC materials with ML.

## 2.2 Polymer-unit library

The polymer unit identified by *PURS* is collected and added to the polymer-unit library. The uniqueness of the polymer unit is ensured, and it is classified by a Python script (**Figure 3a, SI: 1. Polymer-unit SMILES** and **2. Polymer-unit types**). First, the units can be divided into two categories: branch chain and ring units (**Figure 3b**). Branch chains are classified by the elements they contain (**Figure 3c**). Ring units are first classified by the number of rings and then by the element. Through this classification approach, the polymer-unit grouping rules of materials with high mobility can be analysed. Moreover, new materials can be obtained by combining different types of polymer units.

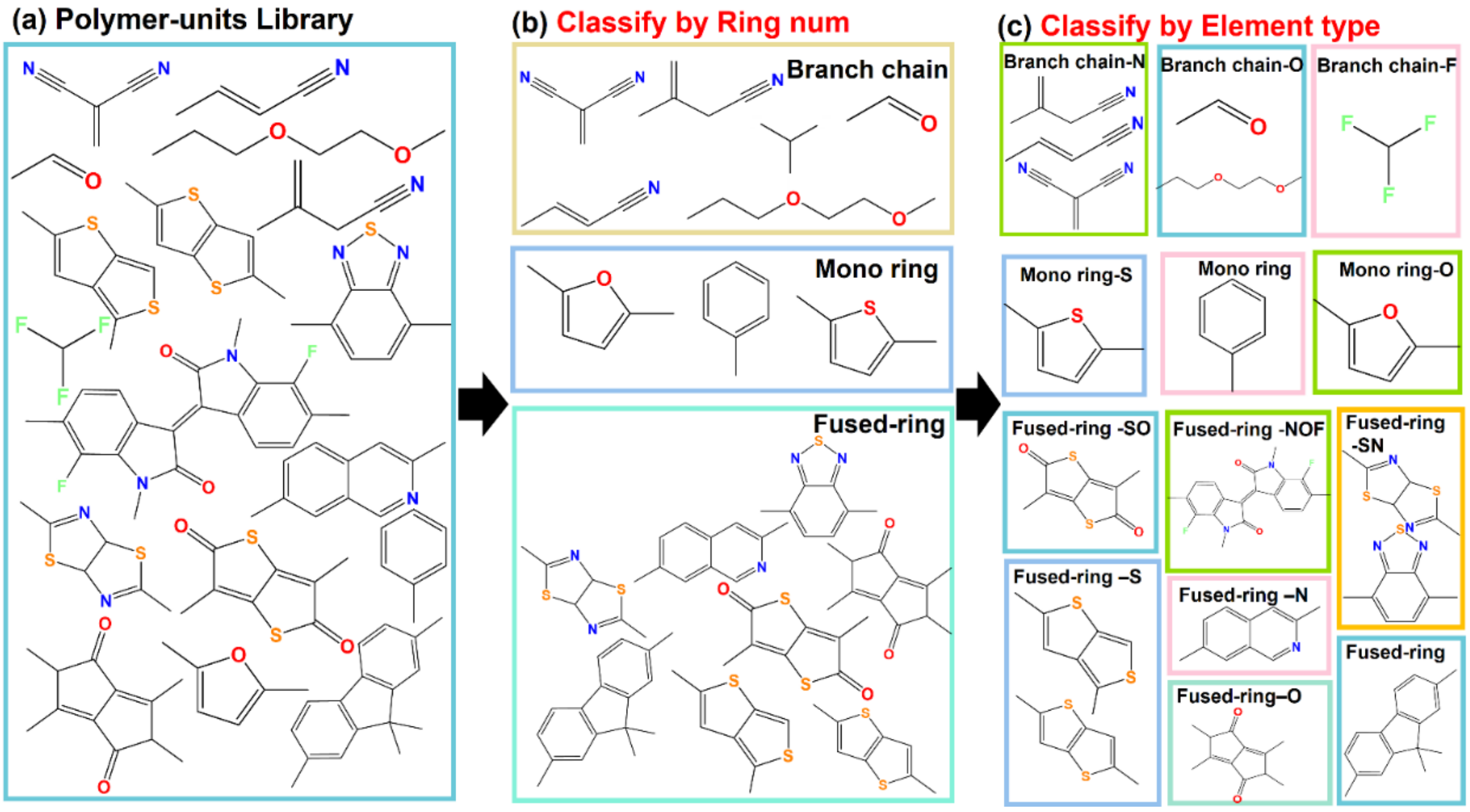

**Figure 3**. **Polymer-unit library.** Polymer units are classified according to the number of rings and the type of elements.

### 2.3. Procedure for acquiring the polymer OSC database & generation strategy for a polymer-unit fingerprint (*PUFp*)

The polymer OSC database contains 678 SMILES codes (standardized by RDKIT [45]) of the polymer OSC materials obtained from published references [7-8, 10-11, 13, 17, 46-57], and each data point was labelled with *N* or *P* carrier mobility labels ($cm^2 \cdot V^{-1} \cdot s^{-1}$ as the unit) (**Figure 2a**). Here for clarity, it should be stated that whether a polymer is *N* or *P* type depends on the electrode identity as well as the chemistry, and the collected experimental data is hard to standardized to all have the same electrode structures. Although the data from different literatures have different experimental conditions, there must be some meaningful correlation between the size of the mobility and the structure of the polymer. The distribution of N-type and P-type data and its influence are discussed in Fig.S2-S3. Polymer units are considered the basic functional building blocks for polymer molecular structure construction. An in-house designed script called *PURS* was prepared to recognize polymer units and generate corresponding fingerprints according to the standardized SMILES codes. The *PURS* program can identify and divide all the polymer units in the SMILES of each OSC. *PURS* identifies and divides polymer units from the input SMILES codes following these rules:

1. By dividing molecules at their breakpoints, polymer OSC molecules are separated into several functional polymer units.

2. The breakpoint must be a single bond that connects two independent units (mono ring, bicyclic ring, fused ring and branched chain units[1]).

[1] The carbon chain skeleton of a polymer monomer is the first one connected, called the main chain, and the carbon chain outside the main chain is the branch chain.

3. The divided linear segments that are acyclic structures are uniformly referred to as chains, and if the segment contains less than 5 atoms in a chain, these chain segments should be combined with the neighbouring ring unit to form a new larger unit. Since there are few acyclic segments with lengths that are greater than 5 atoms on the main chain of an organic semiconductor, the other long-chain unit can be considered a branched chain.

4. All polymer-unit SMILESs need to be canonicalized by *RDKIT* to ensure that the same polymer unit corresponds to the same SMILES code.

After all polymer units are collected and repeated data are removed, a complete "polymer-unit library" is summarized (**Figure 2c**). Here, if the number of polymer units in the polymer-unit library is T and the maximum number of polymer units in the data are N, then we assume a "node matrix" in which the numbers of rows and columns are *T* and *N*, respectively, exists, and "1"s and "0"s are used to describe the presence or absence of particular polymer units in an OSC molecule. Obviously, the node matrix expresses that each OSC data point is composed of polymer units, and the matrix expresses the number of each polymer unit. Finally, each row of the node matrix is summed to generate a one-dimensional vector/fingerprint, named *PUFp*. In this work, the 678 polymer OSC SMILESs in the polymer OSC database (**Figure 2a**) were indexed by a polymer-unit library to generate the corresponding *PUFp* model in one-hot form (**Figure 2d**) [58], as *PUFp* contains information about the type of polymer units and the number of each polymer unit. The length of the generated fingerprint is 445 bits, which means that the 687 OSC data points in this work are composed of 445 different polymer units.

**2.4 ML algorithms & statistical analysis based on *PUFp***

With *PUFp* as input, five ML algorithms, the RF [59], K-nearest neighbour (KNN) [60-61], Gaussian process [62], support vector machine (SVM) [63], and multilayer perceptron (MLP) models [64], are implemented with Python scripts using *scikit-learn* [65] to test the classification performance of different algorithms (**Figure 2e**), First, the *N/P* classification of polymer OSC data is performed by the ML models. *n*- or *p*-type OSCs are defined as follows: if the electron mobility is greater than the hole mobility, the OSC is considered to be an *n*-type OSC; otherwise, it is a *p*-type OSC. Then, according to the experimental mobility values in the polymer OSC database, the mobility classification models for *n*- or *p*-type polymer OSCs are trained and tested by the five ML algorithms mentioned above. The mobility is classified as high (> 4 $cm^2 \cdot V^{-1} \cdot s^{-1}$), medium (1 to 4 $cm^2 \cdot V^{-1} \cdot s^{-1}$), or low (0 to 1 $cm^2 \cdot V^{-1} \cdot s^{-1}$). According to the above classification, there are 477 *p*-type semiconductors, 207 n-type semiconductors, 107 high mobility data, 212 medium mobility data, and 500 low mobility data the polymer OSC database.

*PUFp* can not only be used as input for ML but also be used to perform statistical analysis on the database (**Figure 3**). The types and number of polymer units in the database are analysed, and the polymer-unit combinations of OSCs with high mobility are explored. The best ML model among the five algorithms is used to screen the new materials (See Fig.S1 for more machine learning results on PURS).We further predict the potential mobility values of the newly designed polymer OSC materials based on the combination of the selected key basic polymer units that greatly and positively influence the OSC's mobility performance. Compared to the published materials with the same type of polymer units, the rationality of these high-mobility polymer-unit combinations is verified.

## 3. Results and discussion

### 3.1 Reliability of *PUFp* as a structural expression of the polymer OSC molecule in ML modelling.

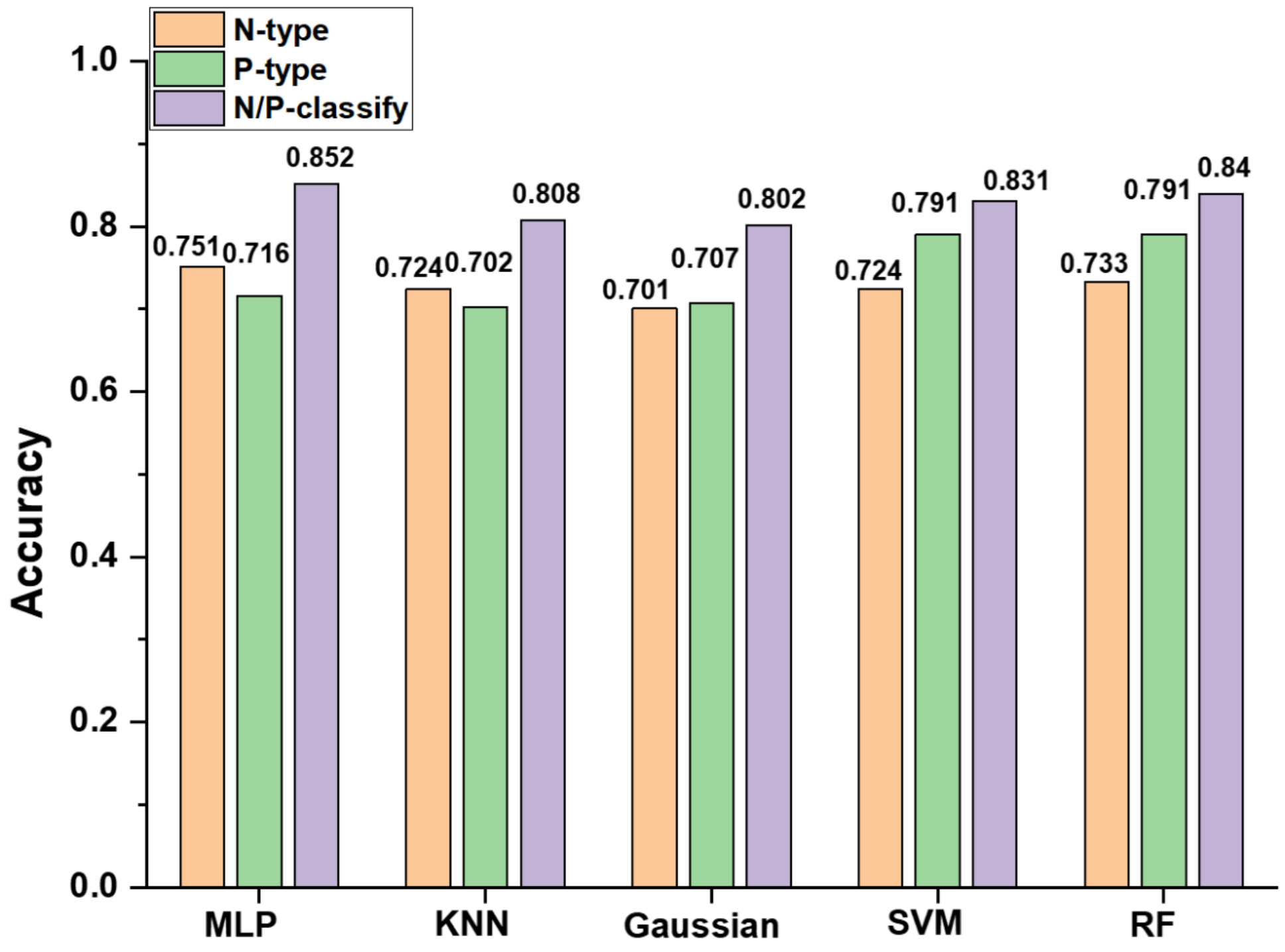

**Figure 4.** Performance of the ML models (MLP, KNN, Gaussian, SVM and RF) using *PUFp* as input.

The *PUFp* model (445 bits in this work) is used as input to train and test the classification performance of the *MLP*, *KNN*, *Gaussian*, *SVM* and *RF* models. The *GridSearchCV* is used to adjust the parameters of each algorithm, and *10-fold cross-validation* are performed, see **Table S1-Table S3**. As shown in **Figure 4**, the N/P classification accuracy of all five algorithms can reach more than 80%. Then, based on the database, the mobility of *n*- or *p*-type OSCs is classified into three categories: high (> 4 $cm^2 \cdot V^{-1} \cdot s^{-1}$), medium (1 to 4 $cm^2 \cdot V^{-1} \cdot s^{-1}$) or low (0 to 1 $cm^2 \cdot V^{-1} \cdot s^{-1}$). The mobility classification test accuracy of both algorithms for n- and p-type OSCs is more than 70%. We note that among the five algorithms, MLP algorithm achieved the highest prediction accuracy of 75.1% for *n*-type OSCs, while SVM and RF algorithm were both reach 79.1% prediction accuracy for *p*-type OSCs. The highest accuracy of *N/P* classification was 85.2%, which came from MLP, slightly higher than RF algorithm (84.0%).

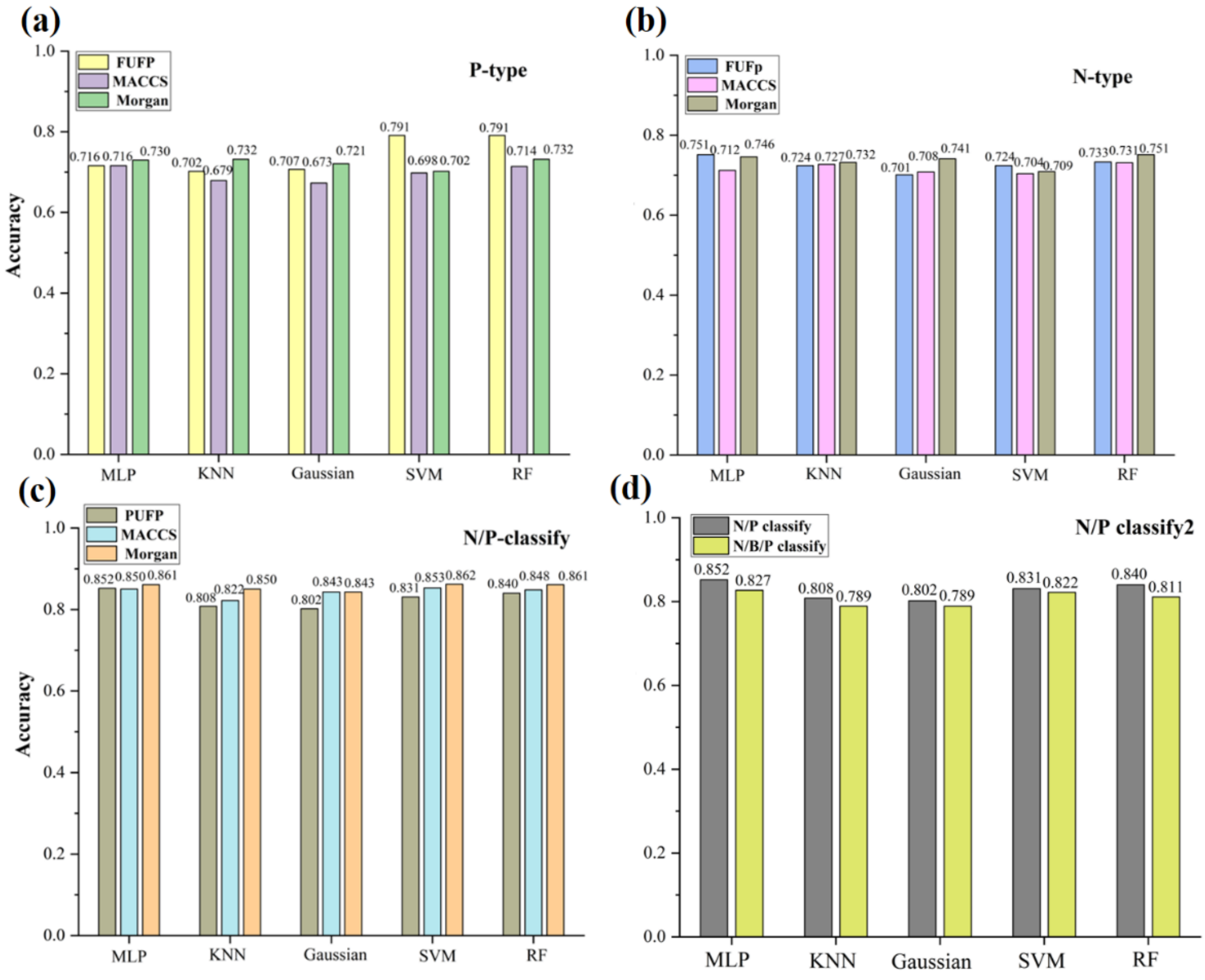

**Figure 5.** (a),(b),(c): Comparison of machine learning accuracy of MACCS,Morgan and FUFp.

(d): Comparison of machine learning accuracy of N/P classify and N/B/P classify.

In addition to FUPp, MACCS(166bits) and Morgan(1023bits) molecular fingerprints are molecular fingerprints with structure for each key. Figure 5 (a-c) is a comparison of these three fingerprints. In general, there was little difference in accuracy between the three fingerprints. In P-type classification, FUFp performs more prominently in SVM and RF algorithms； In N type classification and N/P classification, the best performance is Morgan fingerprint. The ambipolar of OSC is also a topic of considerable concern, so ambipolar is added to the N/P classification (Figure 5-d, the definition of bipolar OSC in this work is: N-type and P-type carrier mobility greater than 1 cm2v-1s-1). It is found that the accuracy of N/B/P classification is slightly lower than that of N/P classification, but it still has 82.7% accuracy under MLP algorithm.

From the results described above, we can conclude that, generally, these algorithms are based on different rules of statistical analysis but can establish accurate relationships between the chemical structure and mobility when *PUFp* is used as input for building ML models to predict mobility because *PUFp* is a notable structural expression of polymer molecules that is easily accessible and contains important chemical information (the type of polymer units and the number of each polymer unit of the polymer molecule).

**3.2 Identification of key polymer units that greatly influence OSC mobility by ML.**

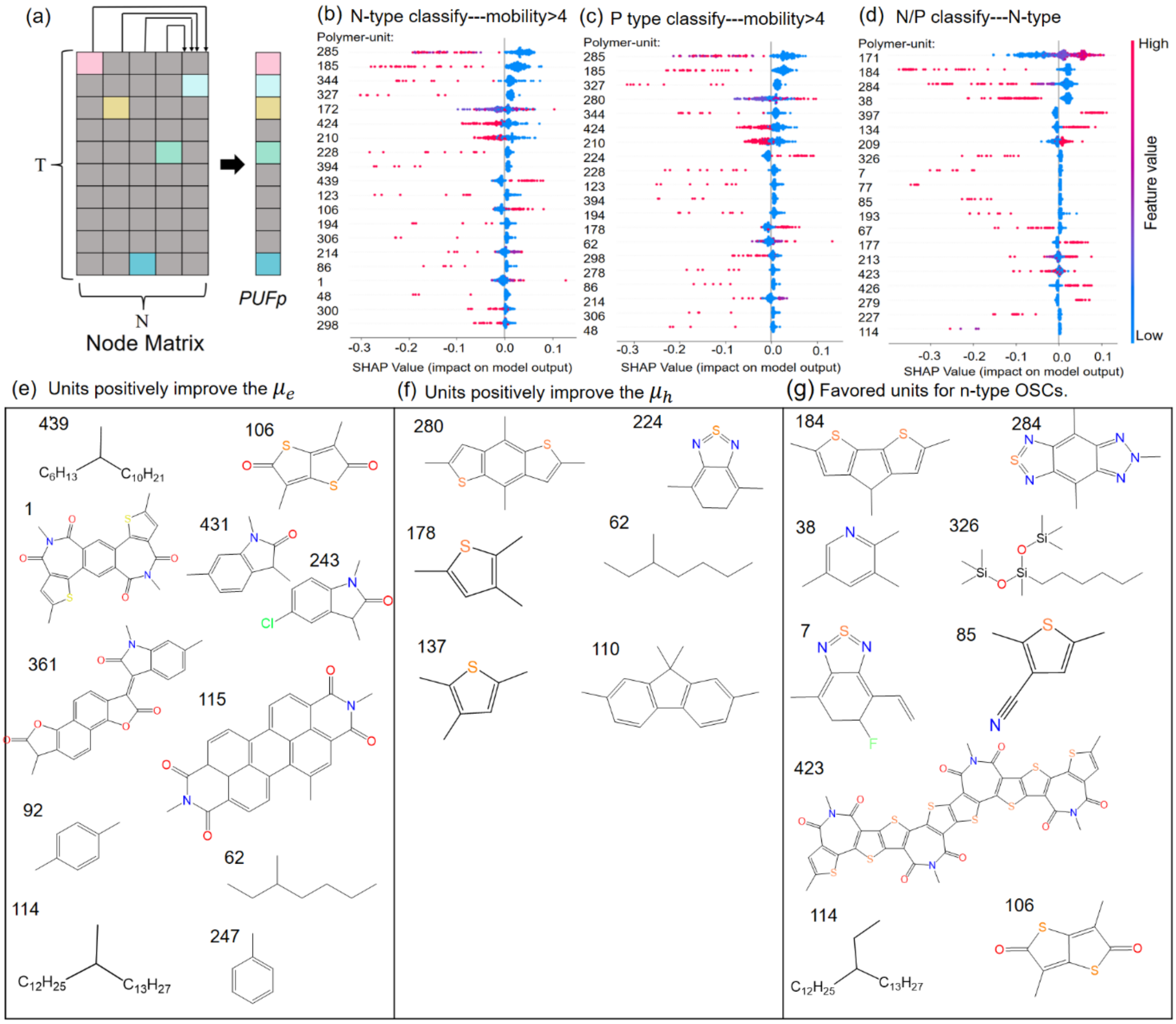

**Figure 6. Identification of key polymer units by using a *SHAP* evaluation**. (a) The generation strategy of *PUFp*; (b, c, d) The interpretations of the ML models for *n*-type, *p*-type and *N/P* classification by the *SHAP* evaluation. The blue and red bars on the right denote the proportional relation between the units and the prediction values. (e, f, g) Chemical structures of the polymer units and their roles are identified through the importance analysis.

The *RF* method performs the well both in *N/P*-type classification and mobility prediction of polymer OSCs (**Figure 4**) because its strategy is to choose multiple features rather than all features from the input for establishing a relationship [66], which is advantageous when dealing with complex and long input. Therefore, we performed a *SHAP* evaluation based on the *RF* model to identify the key polymer units. *SHAP* is a feature importance analysis method based on a tree model (the *RF* model is used for this work) [43]. The *SHAP* value can be positive or negative and reflects the contribution of the corresponding feature to the specified category of data. The *SHAP* evaluation was carried out for n- or p-type high-, medium- and low-mobility classification and *N/P* classification, respectively, with *PUFp* as input. The polymer-unit features (**Figure 6 b, c, d**) with relatively large *SHAP* values in the three *RF* models (*n*-type, *p*-type, *N/P* classification) were analysed and labelled as important or favoured polymer units.

OSCs constructed by favoured polymer units are more likely to be classified into a specified category (for example, OSC structures with favoured polymer units positive for n-type mobility are more likely to be classified as *n*-type high-mobility OSCs). For *n*-type OSCs, the favoured polymer units are listed in **Figure**

**6e**. The serial No. refers to its index number in the polymer-unit library (**Figure 6c**). The *No. 1* and *No. 115* polymer units contain imide groups, the electron-withdrawing groups that contribute to lowering the LUMO level of the polymer and facilitating electron injection into the conduction band [67]. The main function group of the *No. 431, No. 243, and No. 361* polymer units is isoindigo [68]. The structure of isoindigo has good plane properties and a strong electron absorption ability, isoindigo is always used as an electronic Acceptor in *D-A* or *A-A* polymer OSCs, and isoindigo favours a reduction in the *π-π* stack distance.

For *p*-type OSCs, the favoured polymer units are listed in **Figure 6f**. Polymer unit *No. 110* is a full carbon fused ring with large conjugate π bonds, and the complanate delocalized orbitals are beneficial to carrier transport. Polymer unit *No. 224* contains a thiazole structure [69], and the electrostatic attraction between the sulphur and nitrogen atoms in thiazoles is beneficial to forming a closer *π-π* packing structure, which is a common strategy in the *D-A* OSC design [70-71].

**Figure 6g** shows the favoured polymer units for *n*-type OSCs in the *N/P* classification task. Here, *No. 284* and *No. 7* are both polymer units with thiazole structures. This result is in agreement with the results in previous work, in which the low reorganization energy of thiazole can promote the hopping of electrons between units in OSC molecules [53], and the electron affinity of thiazole units is always higher than 3 eV and can reduce the barrier for electron injection [72-75]. Polymer unit *No. 423* is a *BTI* structure and is an effective accepter polymer unit in *n*-type OSCs. Unit *No. 38* with nitrogen, unit *No. 85* with cyanide and unit *No. 106* with aldehyde are all modified electron-withdrawing groups, and their electron-deficient effect is beneficial for improving the air stability of *n*-type semiconductors [76-78]. In addition, the remaining polymer units in **Figure 6e** to g (*No. 431, No. 243, No. 361, No. 178, No. 137, No. 280, No. 110, and No. 224*) are commonly used π-spacer or side-chain structures in OSC molecules.

### 3.3 Exploration of favoured polymer-unit combinations for the mobility of OSCs

To explore the relevance of polymer-unit combinations to OSC mobility, we first classify the polymer units into three groups (mono ring, bicyclic ring or fused ring, and branched chain groups) according to the backbones that frequently appeared in high-mobility OSCs, i.e., *D-A*, *D-D*, and *A-A* with branched chains. The classification and abbreviation rules for the polymer units are defined in **Figure 7c** and are named according to the polymer-unit groups (mono ring, bicyclic ring or fused ring, and branched chain groups) and the element modification approach (the formula of modification elements). Then, the combinations of the mono ring, bicyclic ring or fused ring, and branched chain groups in each OSC of the database were counted by statistically analysing the *PUFp* models. There were 19 kinds of branched-chain combinations, 28 kinds of mono-ring combinations and 75 kinds of bicyclic- or fused-ring combinations. For the details of each combination category, please see SI (SI 2.3 Fused/bicyclic ring combination). A ternary map algorithm was performed and the mono ring, bicyclic ring or fused ring, and branched chain categories were used as the axes for all OSCs in the studied database; see **Figure 7a** and **Figure 7b**. It is found that the OSCs in the studied database are concentrated around several certain combinations. The ternary map has many empty areas in both the *n*- and *p*-type OSCs, and obviously, these unreported combinations generate a huge number of structures (19×28×75=39900). To reduce the number of unreported candidates OSC materials, the scope of the mono ring, bicyclic ring or fused ring, and branched chain group is limited to polymers with high mobility. A scheme for high-mobility combination prediction with ML is shown in **Figure 7d**. The OSC monomer consists of a mono ring, bicyclic ring or fused ring, and branched chain. The Cartesian product of the mono ring-type combination, bicyclic ring or fused ring-type combination and branched chain-type combination in the search space is taken to form a polymer-unit type combination of OSCs. Each combination generates 500 pieces of data. The carrier mobility of these data is predicted using the trained *RF* model shown in **Figure 4**. The ratio of predicted high-mobility data among 500 data points for each combination was calculated, and the combinations with the highest ratios were found.

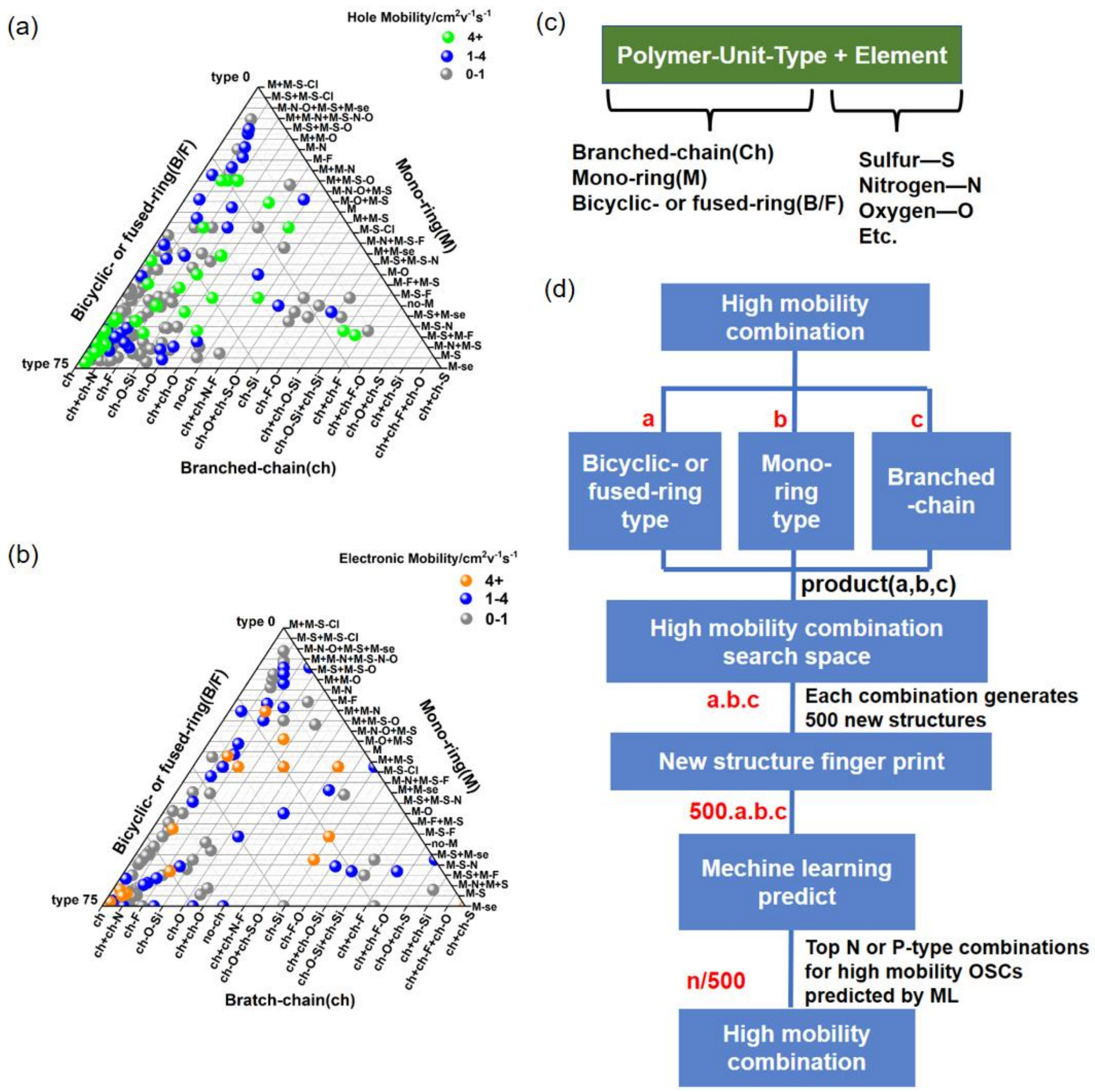

**Figure 7.** Exploring favoured polymer-unit combinations for the mobility of OSCs. (a, b): Ternary distribution of mobility with different combinations of polymer units for *n*- and *p*-type OSCs in the studied database, with the categories of mono ring, bicyclic ring or fused ring, and branched chain as axes. (c) Classification and abbreviation definitions for the polymer units. (d) Scheme of high-mobility combination prediction by ML.

According to the high-mobility combination predicted with ML shown in **Figure 8d**, the favoured polymer-unit combinations for high-mobility OSCs and the top five *n*-type and *p*-type polymer-unit combinations with high-mobility OSCs predicted with ML are listed in **Figure 8l** and **Figure 8m**, respectively. **Figures 8a** to **8k** show the high-mobility polymer OSC structures ($\mu > 4$ $cm^2 \cdot V^{-1} \cdot s^{-1}$) reported in published works, and in these polymer structures, one or more polymer units are of the same type as those in the polymer-unit combination shown in **Figure 8l** and **Figure 8m**. The boundary colours of the polymer-unit combination types in **Figure 8l** and **Figure 8m** are set the same as the colour of the corresponding polymer-unit structure in **Figure 8a** to **8k**.

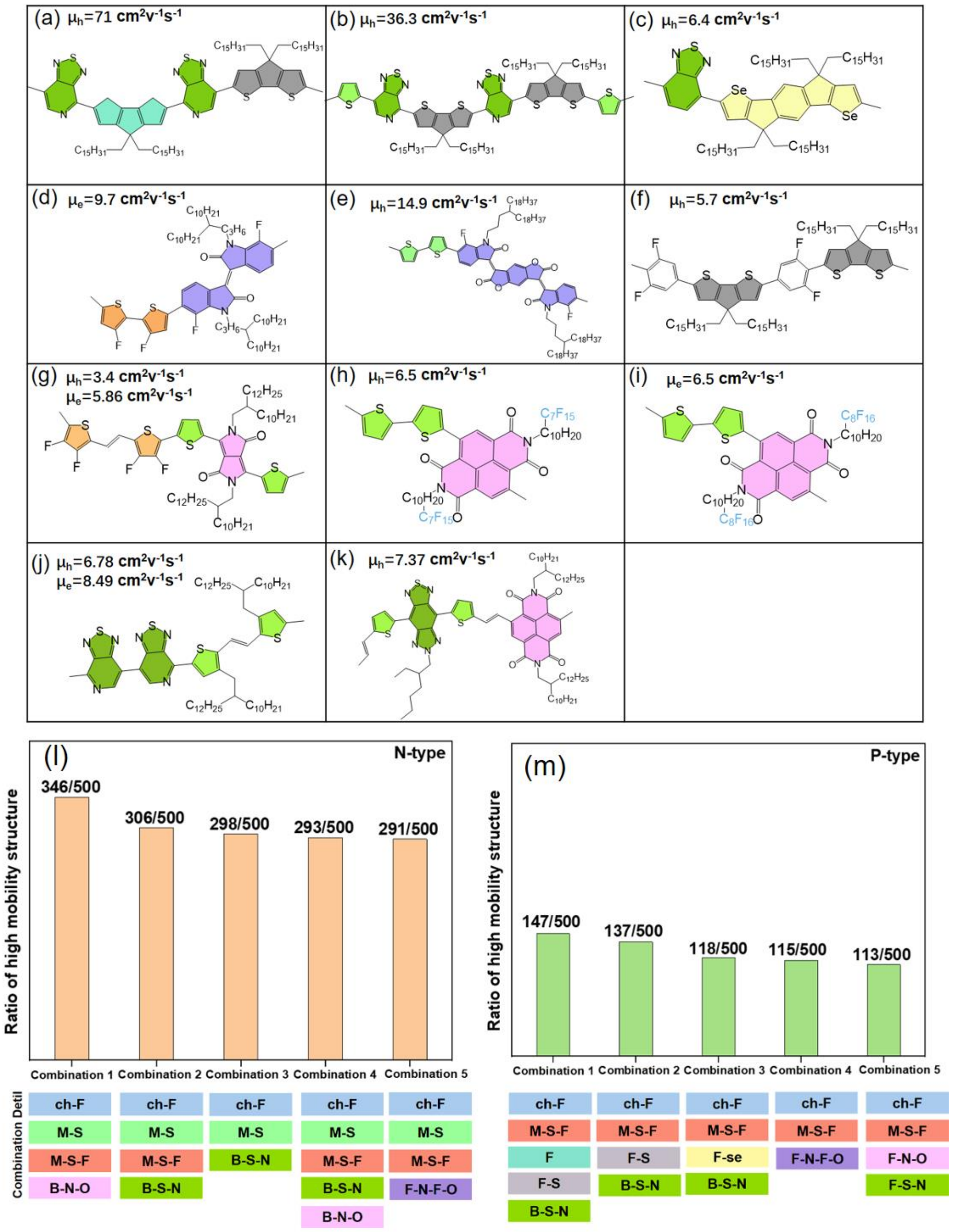

**Figure 8. Favoured polymer-unit combinations with high-mobility OSCs.** (a-k): Sample high-mobility OSC molecules of the same type of polymer-unit combinations [18, 47, 50, 53]. (l, m): Top five *n*-type and *p*-type polymer-unit combinations with high mobility OSCs predicted with ML.

**(这部分内容移至结论部分)**As shown in **Figure 7l** and **Figure 7m**, among the top five *n*-type and *p*-type polymer-unit combinations of high-mobility OSCs predicted by ML, eight combinations are multiple-D-A combinations, indicating that the synergy between multiple *D-A* polymer units is advantageous for mobility [79-82]. Due to the large dipole moment between *D-A* polymer units, such as *isopindus* [11] or *DPP* structures [14] as acceptor units and *thiophene* as donor units, the $\pi$-$\pi$ stacking [50, 53] distance between polymer molecular layers is small (**Figure 7d** and **g**), and the mobility is improved. Moreover, the dihedral angles between adjacent polymer units also affect the carrier mobility [13], and improving the coplanar properties between polymer units is the key to increasing the mobility. Selecting suitable polymer units can improve the coplanar properties. For example, the combination of *thiazole* (acceptor) and *CDT* (cyclopentadithiophene, donor, **Figure 7a, b**) or *IDSe* (indacenodiselenophene, **Figure 7c**) has good coplanar properties [11]. Figure 7-l,m summarize the top five Polymer-Units combinations with high mobility ratio of type N and type P. In these combinations, all have fluorinated side-chain. In addition, 9 of the 10 combinations have fluorinated substitution structures such as single ring containing fluorine. It fully illustrates the importance of fluorine substitution in improving mobility.

Close intermolecular packing is an important factor for the efficient charge transport of OSCs. Branched chain engineering, such as adjusting the length, shape, and modification approach, can not only increase the solubility of an OSC polymer but can also effectively reduce the $\pi$–$\pi$ stacking distance. Semi or fully fluorinated branch chains can strengthen the rigidity and improve the self-assembly effect of OSC materials [50]. As shown in **Figure 7l, m**, all top polymer-unit combinations of high-mobility OSCs have fluorinated branched chains. This is because fluorinated branched chains (ch-F) can improve the regularity of crystallization. In addition, most of the combinations contain fluorinated units (M-S-F, F-n-O-F) because using fluorine to modify a polymer unit can increase the intermolecular force and improve the coplanar properties of the OSCs (**Figure 7d, f, g**).

Modulating the electron-donating or electron-withdrawing strength of a repeating unit in *D-A* combinations results in an optimum frontier molecular energy level that can improve the carrier mobility [11]. The molecular energy levels of *D-A* combination OSCs can be tuned by substituting appropriate groups in the polymer units with strong electron-withdrawing elements (fluorine, pyridine, nitrile, etc.). This approach lowers the LUMO energy level of the conjugated polymer to improve electron transport. In addition, substituting appropriate groups in the polymer units with strong electron-withdrawing elements (such as fluorine) can change the shapes of HOMO and LUMO along the conjugated backbone of OSC, and the appropriate unit modifications and combinations can improve the intermolecular packing and wavefunction overlap, which also benefit the mobility (**Figure 7 d, e, f, g**)[13]. The 10 combinations of Figure 7 (l, m) are analyzed. Each combination consists of three components: branched chain, mono-ring, bicyclic- or fused-ring. Among them, branched chains are used to regulate polymer stacking structure, mono-rings are mostly used as $\pi$ bridges, and bicyclic- or fused-rings are electron donors or acceptors. By comparing N-type and P-type combinations (Figure 7 l, m), it is found that the single rings in P-type combinations are fluorinated sulfur-containing single rings (i.e. fluorothiophene). Among the N-type combinations, the polymer units consists of both sulfur mono-ring (thiophene) and fluoro-sulfur mono-ring (fluorinated thiophene). In other words, the proportion of fluorothiophene can be higher in P-type combinations, because P-type polymer semiconductors usually require a low HOMO level (-5 eV)[83], and fluorine plays a role in lowering the HOMO level.

**4. Conclusion:**

In summary, based on a database containing realistic polymer OSC materials collected from the literature, an in-house designed script called *PURS* was demonstrated to recognize polymer units and generate corresponding fingerprints with the *PUFp* approach, which is a newly developed programming language for expressing OSC molecules and provides distinct results and easy access. A polymer-unit library that consists of 445 units is constructed. Through *SHAP*, the key polymer units for the mobility of OSCs have been identified, and the top-positive-important units for *N*- or *P*-type mobility and *N/P* classification are discovered.

The purpose of this study is to promote PURS program, and use the PUFP generated by PURS combined with data analysis method to analyze the mobility of organic polymer semiconductors.

The PURS program combined with data analysis methods can achieve:**(Reviewer 2 Q13)**

1. By using PURS, each polymer-unit in a Polymer monomer can be identified and a polymer-Units database can be created.

2. PURS can be generated by PURS program. Using PUFp as the input feature, the trained machine learning model can achieve more than 80% accuracy in the polymer semiconductor database in this paper. PURS are therefore practical molecular fingerprints. And because each key of the PUFp represents a specific polymer-unit, it is highly interpretable.

3. The machine learning results of PUFp were analyzed using SHAP to identify the polymer-unit that is conducive to the increase of P-type and N-type mobility, and the polymer-unit that is conducive to the formation of N-type polymer semiconductors.

4. 10 combinations of high mobility are summarized. The high mobility polymer search space is obtained by analyzing the existing high mobility polymer combination types. PUFp were generated for the structures in this space, and the ratio of structures predicted to be highly mobile among these structures was counted (using the model trained in 2).

We developed a scheme to help polymer OSC material design by combining ML approaches and PUFp as input not only to passively predict OSC mobility but also to actively provide structural guidance for new high-mobility OSC material design. That is, a large number of polymer OSC materials could be screened through a pre-evaluation and classification process by using our methods, and then the identified leading candidates could be synthesized and further tested by experiments.

Overall, our framework for *PURS* generation and its application with ML is verified to possess the ability to design high-mobility polymer OSC materials. It is believed that when coupled with the Materials Genome

Initiative and ML, this framework will bring even greater changes to high-mobility polymer OSC material discovery.

## Acknowledgements

**Funding:** Financial support was provided by the Natural Science Foundation of China (92163212), Fundamental Research Program of Shenzhen (JCYJ20190809174203802), Guangdong Provincial Key Laboratory of Computational Science and Material Design (2019B030301001), Guangdong Provincial Young Innovative Talents (2019KQNCX136) and the Key Research Project of Zhejiang Lab (2021PE0AC02). W. Zhang acknowledges support from the Guangdong Innovation Research Team Project (2017ZT07C062). Computing resources were supported by the Center for Computational Science and Engineering at Southern University of Science and Technology.

**Author contributions:** C. Ye and W. Zhang formulated this project. X. Zhang performed data collection, program coding, and ML analysis. G. Wei and Y. Sheng provided helpful discussion for program coding. X. Zhang and C. Ye cowrote the manuscript. W. Zhang revised the manuscript. W. Zhang and C. Ye secured the funding.

**Competing interests:** None declared.

**Data and materials availability:** All data and model analyses are available in the supplementary materials.

# 1.Polymer-units smiles

(1) Bratch-chain
11.‘CCCCCCC(CC)CCCCCC',
27.'CCCCCCCCCCCCCCCCCCCCCCC',
29.'CCCCCCCCCCCCCCC(CCCCCC)CCCCCCCCCCCC',
31.'CCCCCCCCCCCCC(CCCCCCC)CCCCCCCCCC',
39.'CCCCCCCCCCC(C)CCCCCCCC',
42.'CCCCCCCCCCCCC(CCC)CCCCCCCCCC',
46.'CCCCCCCCCCCCC(CCCCC)CCCCCCCCCC',
51.'CCCCCCCCCCC(CC)CCCCCCCC',
56.'CCCCCCC(CCCC)CCCC',
62.'CCCCC(C)CC',
65.'CCCCCCCCCCCCCCC(C)CCCCCCCCCC',
68.'CCCCCCCCC(C)CCCCCC',
69.'CCCCCCCCC',
75.'CCCCCCCCCCC(CCC)CCCCCC',
87.'CCCCCCCCCCCCCCCCCCC',
109.'CCCCCCCCCCC(CCC)CCCCCCCCCC',
114.'CCCCCCCCCCCCC(C)CCCCCCCCCCCC',
144.'CCC(C)CCCC(C)C',
147.'CCCCCCC(C)CCCCCC',
158.'CCCCCCCCCCCCCCC(C)CCCCCCCCCCCC',
163.'CCCCCCCCCCCCCCC(CCCC)CCCCCCCCCCCCCC',
167.'CCCCCCCCCCC(CCCC)CCCCCCCCCC',
171.'CCCCCCCCCCC(CCCC)CCCCCCCC',
181.'CCCCCCC(CCCC)CCCCCC',
191.'CCCCCCCCCCCCCCC(CCC)CCCCCCCCCCCCCC',
193.'CCCCCCC(C)CC',
195.'CCCCCCCCCCCCCCCCCCCCC',
196.'CCCCCCC(CCCC)CCCCC',
206.'CCCCCCC(CCC)CCCC',
213.'CCCCCCCCCCCCC(CC)CCCCCCCCCC',
214.'CCCCCCCCCCCC',
216.'CCCCCCCCCCCCC(CCCC)CCCCCCCCCC',
221.'CCCCCCCCCCCCCCC',
231.'CCCCCCCCCCCCC',
238.'CCCCCCCCCCCCC(C)CCCCCCCC',
250.'CCCCCCC',
252.'CCCCCCCC',
257.'CCCCC',
258.'CCCCCC',
265.'CCCCCCC=CCCCCCCCC',
267.'CCCCCCCCCCCCCCCCCCCCCCCCCCCCCCCCCCCCC',
270.'CCCCC(CC)CCC',
276.'CCCCCCC(CCC)CCCCCC',
283.'CCCCCCCCCCCCCCCCC',
285.'CCCCCCCCCCCCC(C)CCCCCCCCCC',

298.'CCCCCCCCCCCCCC',
300.'CCCCCCCCCC',
304.'CCCCCCCCCCC(CC)CCCCCCCCCC',
310.'CCCCCCCCCCC(C)CCCCCCCCCC',
315.'CCCCCCCCCCCCCCC(C)CCCCCCCCCCCCCC',
320.'CCCCCCCCCCCCCCCCC(C)CCCCCCCCCCCCCC',
327.'CCCCCCCCCCCCCCCCCCC(CCC)CCCCCCCCCCCCCCCCCC',
329.'CCCCCCCCCCCCCCCCCCCC',
333.'CCCCC(CC)CCCC',
334.'CCCCCCCCCCC(CCCC)CCCCCCCCC',
335.'CC(C)CCCC(C)C',
337.'CCCCCCCCC(C)CCCCCCCC',
346.'CC(C)C',
352.'CCCCCCCCCCCCCCCCCCC(CCCC)CCCCCCCCCCCCCCCCCC',
355.'CCCCCCC(C)CCCC',
357.'CCCCCCCCCCCCCCCCCC',
366.'CCCCCCCCCCCCCCCCCCC(C)CCCCCCCCCCCCCCCCCC',
368.'CCCCCCCCCCCCCCCCCCC(CC)CCCCCCCCCCCCCCCCCC',
373.'CCCCCCCCCCCCC(CCCCCCCCC)CCCCCCCCCC',
375.'CCCCCCCCCCC(CCC)CCCCCCCC',
380.'CCCCCCCCCCCCC(CCCCCC)CCCCCCCCCC',
382.'CCCCCCCCCCCCC(CCCCCCCC)CCCCCCCCCC',
390.'CCCCCCCCCCC',
396.'CCCCCCCCCCCCCCCCCCC(CCCCCC)CCCCCCCCCCCCCCCCCC',
397.'CCCCCCCCCCCCCCC(CCC)CCCCCCCCCC',
398.'CCCCCCCCCCCCCCCC',
404.'CCCCCCCCCCC(CCCCCC)CCCCCCCCCC',
413.'CCCCC(CC)CC',
428.'CCCCCCCCCCCC(C)CCCCCCCCCC',
432.'CCCCCCCCCCCC(CCCC)CCCCCC',
433.'CCCCCCCCCCCCCCCCCCCCCC',
439.'CCCCCCCCCCC(C)CCCCCC',
444.'CCCCCCCCCCCCCCCCCCC(CCCCC)CCCCCCCCCCCCCCCCCC'

(2) Chain-F

94.'CCCCCC(F)(F)C(F)(F)C(F)(F)C(F)(F)F',
253.'FC(F)F',
287.'CCCCCCCCCCC(F)(F)C(F)(F)C(F)(F)C(F)(F)C(F)(F)C(F)(F)C(F)(F)C(F)(F)F',
293.'CCCCCCCCCCC(F)(F)C(F)(F)C(F)(F)C(F)(F)C(F)(F)C(F)(F)C(F)(F)F',
318.'CCCCCC(F)(F)C(F)(F)C(F)(F)C(F)(F)C(F)(F)C(F)(F)C(F)(F)F',
370.'CCCCCC(F)(F)C(F)(F)C(F)(F)C(F)(F)C(F)(F)C(F)(F)F',
387. 'CCCCCC(F)(F)C(F)(F)C(F)(F)C(F)(F)C(F)(F)C(F)(F)C(F)(F)C(F)(F)F',
435.'CCCCCCCCCC(F)(F)C(F)(F)C(F)(F)C(F)(F)C(F)(F)C(F)(F)C(F)(F)F'

(3) Chain-F-O

197.'CCCCCOCCC(F)(F)C(F)(F)C(F)(F)C(F)(F)C(F)(F)C(F)(F)F',
294.'CCCCCOCCC(F)(F)C(F)(F)C(F)(F)C(F)(F)C(F)(F)C(F)(F)C(F)F’

(4) Chain-N

129.'CCC#N',
146.'CCCCCCCCCCCCCCN',
159.'CCCCCCCCCCN',

248.'C=C(C)C#N',
305.'CC=CC#N',
311.'N#CCC#N',
313.'C=C(C#N)C#N

(5) Chain-N-F

43.'CCCCCCCCCCC(CCCCCCCC)CNB(F)F'
369. 'NB(F)F'

(6) Chain-O

2.'CCCCCCCCCCCCCCCCC(CCCCCCCCCCCCCCCC)OC(=O)CCC',
6.'CCCCCCC(CCCC)OC=O',
9.'CCCCCCCCCCO',
15.'CCCOCCOC',
21.'COCCOC',
23.'CCCCCCCCOCCOCCOCCCCO',
25.'CCCCCCCCCCCCOC(=O)CCC',
26.'CCCCCCC(CO)CCCC',
55.'CCCCCCCCOCCOCCOCCOC',
57.'CCCCCCCCCCCCO',
59.'CC=O',
71.'CCCCCCC(O)CCCC',
74.'COCCOCCOCCOC',
77.'CCCCCCCCO',
84.'CCOCOCOC',
96.'CCCCCCCCCCCCCCOC(=O)CCCCC',
97.'CCCOCOCCO',
105.'CCCCCCCCCCC(CCCO)CCCCCCCC',
111.'C#CC(C)CO',
113.'CCCCCCCCCCCCOC(=O)CCCCC',
120.'CCCCCCCOCCOCCOCCOC',
125.'CCCCCCCCCCCCCCCCO',
134.'CCCCCCCCCCCCCCOC(=O)CCC',
139.'CCCCCCCCCCCC=O',
157.'CCCCCCCCCCCCOC=O',
166.'CCCCCCCOCCOCCOCCCCO',
174.'CCCCCCCCCCC(C=O)CCCCCCCCCC',
192.'CCCCCCCCCCCCC(CCO)CCCCCCCCCC',
208.'CCCCC(CC)COC=O',
215.'CCOCCOCCOCCOC',
217.'CCCCCCCOC=O',
225.'CCCCCCCCCCC(O)CCCCCCCC',
227.'CC(C)CCCC(C)COC=O',
233.'CCCCCCO',
241.'CCCCCCCCCCCCC(CCCO)CCCCCCCCCC',
255.'CCOCCOCCOC',
260.'CCCCCCCCCCCCCCCCO',
261.'CCCCCCCCCCCCC(CCCCCCCCCC)COC(=O)CCC',
282.'CCCCC(CC)OC=O',
292.'CCCCCCCCCCC(CCCCCCCC)COC(=O)CCC',
295.'CCCCCCCCCCCCC(CO)CCCCCCCCCC',

314.'CCCCCCCCCOC=O',
317.'CCCCCCCCCCC(CO)CCCCCCCC',
319.'CCCCCCCCCCCCCCCCCO',
321.'CCCCCCCCCCCCCCO',
323.'CC(=CCCO)CCO',
331.'CCCCCCCCCCCCC(CCCCCCCCCC)COC(=O)CCCCC',
336.'CCCCC(O)CC',
363.'CCCCCCCCCOCCOCCOCCO',
364.'CCOCCOC',
377.'CCCCCCCCOCCOCCOCCOCC',
379.'CCCCC(CC)CO',
384.'COCCOCC(C)COCCOC',
395.'CCCCCCOCCCCOCCOCCOC',
411.'CCCCOCCOCCOC',
425.'CCCCCCCCCCC(CCCCCCCC)COC(=O)CCCCC',
434.'CCOCCOCOC',
437.'CCCCCCCCCCCCC(CCCCO)CCCCCCCCCC',
441.'CCCCCCCCOCCCCC'

(7) Chain-O-Si

19.C[Si](C)(C)O[Si](C)(C)O[Si](C)(C)C
73.CCCCCCC[Si](C)(O[Si](C)(C)C)O[Si](C)(C)C
93.CCCCCCCCCCC[Si](OC)OC
118.CCCCCC[Si](C)(O[Si](C)(C)C)O[Si](C)(C)C
119.CCCCCC[Si]O[Si]C
154.C[Si](C)(C)O
173.CCCCCCCCCCC[Si](C)(O[Si](C)(C)C)O[Si](C)(C)C
244.CCCCCCCCCCC[Si]O[Si]C
246.CCCCCCCCCC[Si](C)(O[Si](C)(C)C)O[Si](C)(C)C
264.CCCCC[Si]O[Si]C
291.CCCCCCC[Si]O[Si]C
297.CCCCC[Si](C)(O[Si](C)(C)C)O[Si](C)(C)C
326.CCCCCC[Si](C)(O[SiH](C)C)O[Si](C)(C)C
349.CCCC[Si](C)(O[Si](C)(C)C)O[Si](C)(C)C
365.CCCC[Si]O[Si]C
407.CCCCCCCCC[Si](C)(O[Si](C)(C)C)O[Si](C)(C)C
429.CCCCCCCC[Si](C)(O[Si](C)(C)C)O[Si](C)(C)C

(8) Chain-S

130.CCCCCCCCS
143.CCCCC(S)CC
420.CCCCCCCCC(S)CCCCCC

(9) Chain-S-O

180.CC(=O)S

(10) Chain-Si

70.CCCCCC[Si](C)(CCCCCC)CCCCCC
148.CCCCCCCC[Si](C)(CCCCCCCC)CCCCCCCC
150.CCCCCCCC[Si](C)(CCCCCC)CCCCCC
168.C#CC#CC#C[Si](C)(CCCCCC)CCCCCC
219.CCC[Si](CCC)CCC
234.CCCCCC[Si]C

341.CCCCCCCCC[Si](C)(CCCCCCCC)CCCCCCCC
(11) Bicyclic
389.Cc1ccc2ccc(C)cc2c1
430.Cc1ccc2cc(C)ccc2c1
(12) Bicyclic-N
152.Cc1ccc2cc(C)ncc2c1
177.Cc1cccc2nc(C)c(C)nc12
205.CC1=C2CN(C)C(C)=C2CN1
240.Cc1ccc2ccncc2c1
249.Cc1cccc2nn(C)nc12
(13) Bicyclic-N-F
40.Cc1c(F)c(F)cc2nn(C)nc12
340.Cc1c(F)ccc2nc(C)c(C)nc12
(14) Bicyclic-N-O
50.Cc1ccc2c(c1)C(C)C(=O)N2
101.Cc1cccc2c1C(=O)N(C)C2=O
201.CC1=C2C(=O)NC(C)=C2C(=O)N1
268.Cc1cccc2nonc12
289.CC1=C2NC(=O)C(C)=C2N(C)C1=O
424.CC1=C2C(=O)N(C)C(C)=C2C(=O)N1
426.CCCCN1C(=O)C2=C(C)N(CCCC)C(=O)C2=C1C
431.Cc1ccc2c(c1)N(C)C(=O)C2C
(15)Bicyclic-N-O-Cl
243.CC1C(=O)Nc2ccc(Cl)cc21
345.Cc1cc2c(cc1Cl)C(C)C(=O)N2C
(16)Bicyclic-O
277.CC1=C2C(=O)C(C)C(C)=C2C(=O)C1
(17) Bicyclic-S
28.Cc1cc2sc(C=C)cc2s1
66.Cc1cc2sc(C)cc2s1
102.Cc1cc2ccsc2s1
108.Cc1sc2ccsc2c1C
123.Cc1cc2sccc2s1
385.Cc1scc2sc(C)cc12
(18)Bicyclic-S-F
199.Cc1scc2c(F)c(C)sc12
(19) Bicyclic-S-N
18.Cc1cncc2nsnc12
24.Cc1ccc(C)c2nsnc12
49.Cc1c(C#N)c(C#N)cc2nsnc12
82.Cc1cc2c(s1)C(C)CN2
95.Cc1nc2csc(C)c2nc1C
116.Cc1scc2c1N=S=N2
128.Cc1nc2scnc2s1
165.Cc1cc2c(s1)C(C)CN2C
169.Cc1scc2nccnc12
224.Cc1cccc2nsnc12
230.Cc1scc2nc(C)c(C)nc12
308.Cc1nccc2nsnc12

330.Cc1cc2nsnc2c(C)c1C
(20) Bicyclic-S-N-F
7.C=Cc1c(F)cc(C)c2nsnc12
122.Cc1c(F)ccc2nsnc12
124.Cc1cc(F)cc2nsnc12
278.Cc1c(F)c(F)cc2nsnc12
(21)Bicyclic-S-N-O
4.Cc1cc2c(s1)C(C)C(=O)N2
190.CC1C(=O)Nc2ccsc21
316.Cc1scc2c1C(=O)N(C)C2=O
338.Cc1cc2c(s1)C(C)C(=O)N2C
(22)Bicyclic-S-O
106.CC1=C2SC(=O)C(C)=C2SC1=O
(23)Fused
12.Cc1cc2c(cc1C)C(C)c1cc3c(cc1-2)C(C)c1cc(C(C)(C)C)c(C)cc1-3
20.Cc1ccc2cc3cc(C)ccc3cc2c1
91.Cc1ccc2c(c1)C(CCCC)(CCCC)c1cc(C)ccc1-2
110.Cc1ccc2c(c1)C(C)(C)c1cc(C)ccc1-2
186.Cc1ccc2c(c1)[Ge](CCCC)(CCCC)c1cc(C)ccc1-2
212.Cc1ccc2c(c1)-c1ccc(C)cc1C2
242.Cc1ccc2c(c1)C(C)(C)c1cc3c(cc1-2)C(C)(C)c1ccccc1-3
290.CC1=CC2=C(C1)C1=C(C=C(C)C1)C2
348.Cc1ccc2c(c1)C(C)c1cc3c(cc1-2)C(C)c1ccccc1-3
401.CC#Cc1c2ccc(C)cc2c(C)c2ccc(C)cc12
436.CC1c2ccccc2-c2ccccc21
438.Cc1ccc2c(c1)C(C)(C)c1ccccc1-2
(24)Fused-N
16.Cc1ccc2c3ccccc3n(C)c2c1
64.Cc1ccc2c3ccc(C)cc3n(C)c2c1
(25)Fused-N-F-O
3.Cc1ccc2c(c1F)N(C)C(=O)C2=C1C(=O)N(C)c2c1ccc(C)c2F
13.Cc1c(F)cc2c(c1F)N(C)C(=O)C2=C1C(=O)Oc2cc3c(cc21)OC(=O)C3=C1C(=O)N(C)c2c(F)cc(F)cc21
117.Cc1ccc2c(c1F)N(C)C(=O)C2=C1C(=O)Oc2cc3c(cc21)OC(=O)C3=C1C(=O)N(C)c2c(F)cccc21
145.CC=Cc1ccc2c(c1F)N(C)C(=O)C2=C1C(=O)Oc2cc3c(cc21)OC(=O)C3=C1C(=O)N(C)c2c1ccc(C)c2F
198.Cc1ccc2c(c1F)N(C)C(=O)C2=C1C(=O)N(C)c2c(F)cccc21
200.Cc1ccc2c(c1F)N(C)C(=O)C2=C1CN(C)c2c1ccc(C)c2F
309.Cc1ccc2c(c1F)N(C)C(=O)C2=C1C(=O)N(C)c2ncccc21
322.Cc1ccc2c(c1F)N(C)C(=O)C2=C1C(=O)N(C)c2nc(C)ccc21
(26) Fused-N-O
10.Cc1cc2c(cc1C)nc1c3ccc4c5c(ccc(c(=O)n21)c35)c(=O)n1c2cc(C)c(C)cc2nc41
14.Cc1c2nc3c(nc2cc2nc4c(nc12)-c1ccc2c5c(ccc-4c15)C(=O)N(C)C2=O)-c1ccc2c4c(ccc-3c14)C(=O)N(C)C2=O
22.Cc1ccc2c(c1)N(C)C(=O)C2=C1C(=O)N(C)c2ccccc21
34.CC=Cc1ccc2c(n1)N(C)C(=O)C2=C1C(=O)Oc2cc3c(cc21)OC(=O)C3=C1C(=O)N(C)c2nc(C)ccc21
37.CC=Cc1ccc2c(c1)N(C)C(=O)C2=C1C(=O)Oc2cc3c(cc21)OC(=O)C3=C1C(=O)N(C)c2cc(C)ccc21
41.Cc1cc2c3c(c(C#C)cc4c3c1C(=O)N(C)C4=O)C(=O)N(C)C2=O
53.Cc1cc2nc3c4ccc5c6n(c(=O)c7ccc(c(=O)n3c2cc1C)c4c75)C(C)C(C)N=6
58.Cc1cc2c3c(ccc4c5c(C)cc6c7c(ccc(c1c34)c75)C(=O)N(C)C6=O)C(=O)NC2=O
76.Cc1ccc2c(n1)N(C)C(=O)C2=C1C(=O)N(C)c2nc(C)ccc21
78.Cc1c(=O)n2c3ccc(C)cc3c3cc(=O)n4c5ccc(C)cc5c1c4c32

115.Cc1cc2c3c(ccc4c5ccc6c7c(ccc(c1c34)c75)C(=O)N(C)C6=O)C(=O)N(C)C2=O
142.Cc1cc2c3c(cc(C=C)c4c3c1C(=O)N(C)C4=O)C(=O)N(C)C2=O
153.Cc1cc2nc3c4ccc5c6c(ccc(c(=O)n3c2cc1C)c46)c(=O)n1c2cc(C)c(C)cc2nc51
175.Cc1ccc2c(c1)N(C)C(=O)C2=C1C(=O)N(C)c2cc3c(cc21)N(C)C(=O)C3=C1C(=O)N(C)c2cc(C)ccc21
185.Cc1cc2c3c(ccc4c3c1C(=O)N(C)C4=O)C(=O)N(C)C2=O
203.Cc1ccc2c(c1)N(C)C(=O)C2=C1C(=O)Oc2cc3c(cc21)OC(=O)C3=C1C(=O)N(C)c2cc(C)ccc21
256.CC#Cc1cc2c3c(ccc4c5c(C)cc6c7c(ccc(c1c34)c75)C(=O)N(C)C6=O)C(=O)N(C)C2=O
279.Cc1ccc2c(n1)N(C)C(=O)C2=C1C(=O)Oc2cc3c(cc21)OC(=O)C3=C1C(=O)N(C)c2nc(C)ccc21
301.CN=C1c2ccc3c4c2c(ccc4c(=O)n2c4cc(C)c(C)cc4nc32)C(=O)N1C
306.Cc1cc2c3c(c(C)cc4c3c1C(=O)N(C)C4=O)C(=O)N(C)C2=O
324.Cc1ccc2c(c1)c1cc(=O)n3c4ccc(C)cc4c4c(C)c(=O)n2c1c43
328.Cc1cc2c3c(c(C=C)cc4c3c1C(=O)N(C)C4=O)C(=O)N(C)C2=O
361.Cc1ccc2c(c1)N(C)C(=O)C2=C1C(=O)Oc2c1ccc1c3c(ccc21)C(C)C(=O)O3
371.Cc1c(=O)n2c3ccccc3c3cc(=O)n4c5ccccc5c1c4c32
403.Cc1cc2c3c(c(C)cc4c3c1C(=O)NC4=O)C(=O)N(C)C2=O
427.Cc1ccc2c(c1)N(C)C(=O)C2=C1C(=O)N(C)c2cc(C)ccc21
(27) Fused-N-O-Cl
354.Cc1ccc2c(c1Cl)N(C)C(=O)C2=C1C(=O)N(C)c2c(Cl)cccc21
(28) Fused-O
127.Cc1ccc2c(c1)C(CO)(CO)c1ccc(C)cc1-2
226.CC1C(=O)Oc2c1ccc1c3c(ccc21)C(C)C(=O)O3
399.Cc1ccc2c(c1)C(=O)c1cc(C)ccc1-2
(29) Fused-S
5.Cc1cc2c(s1)-c1sc(C)cc1C2
44.Cc1sc2c(sc3c2sc2c4sccc4sc23)c1C
61.Cc1sc2c(sc3ccsc32)c1C
98.Cc1cc2c(s1)c1sccc1c1c(C)sc(C)c21
99.Cc1cc2c(s1)-c1sccc1[Ge]2
140.Cc1cc2cc3sccc3cc2s1
155.Cc1cc2sc3c4sccc4sc3c2s1
156.Cc1cc2c(s1)-c1sc(C)cc1[Ge]2
184.Cc1cc2c(s1)-c1sccc1C2
189.Cc1cc2cc3c(ccc4ccsc43)cc2s1
211.Cc1cc2sc3ccsc3c2s1
236.CC1CC2=C(S1)c1sccc1C2
239.Cc1cc2c3sc(C)cc3c3ccsc3c2s1
263.Cc1cc2cc3c(cc2s1)-c1cc2sccc2cc1C3
269.Cc1cc2ccc3c(ccc4ccsc43)c2s1
272.Cc1sc2c(sc3c4sccc4sc23)c1C
280.Cc1cc2c(C)c3sccc3c(C)c2s1
325.Cc1cc2c(s1)C(C)c1ccccc1-2
343.Cc1cc2sc3cc(C#C)sc3c2s1
350.Cc1cc2c3cscc3c3ccsc3c2s1
374.Cc1cc2c(s1)-c1cc3c(cc1C2)-c1sccc1C3
386.Cc1cc2c3ccsc3c3sc(C)cc3c2s1
388.Cc1cc2c(C)c3sc(C)cc3c(C)c2s1
393.Cc1cc2sc3c(c2s1)C(C)(C)c1cc2c(cc1-3)C(C)(C)c1c-2sc2cc(C)sc12
405.Cc1cc2c(s1)c(C)cc1c3ccsc3c(C)cc21
409.Cc1cc2ccc3ccsc3c2s1

416.CC1c2ccccc2-c2ccsc21
(30) Fused-S-N
48.Cc1c2c(cc3nsnc13)N=S=N2
90.Cc1cc2c(s1)-c1sc3c4c(sc3c1CN(C)C2)-c1sc2c3c(sc2c1CN(C)C4)-c1sc2c4c(sc2c1CN(C)C3)-
c1sc2c3c(sc2c1CN(C)C4)-c1sc(C)cc1CN(C)C3
100.Cc1ccc2c(c1)Sc1cc(C)ccc1N2
162.CC(C#N)=C1Sc2c(c3c4c(c5c(c6c4c2CN(C)C6)SC(C)S5)CN(C)C3)S1
183.Cc1scc2nc3c(nc12)-c1cccc2cccc-3c12
204.Cc1cc2c(ccc3nsnc32)c2nsnc12
229.CC1=NC2C(C)c3nsnc3C(C)C2N=C1C
237.Cc1nc2c(C)c3scnc3c(C)c2s1
266.Cc1nc2cc3scnc3cc2s1
271.CC1c2nn(C)nc2C(C)C2N=S=NC12
284.Cc1c2c(cc3nn(C)nc13)N=S=N2
299.Cc1cc2c3nsnc3c3ccsc3c2s1
339.Cc1nc2cc3nsnc3c(C)c2nc1C
406.CC1=C2N=S=NC2=C(C)C2N=S=NC12
410.Cc1cc2c3nc4ccccc4nc3c3ccsc3c2s1
414.Cc1ccc2c(c1)N(C)c1ccc(C)cc1S2
423.Cc1cc2c(s1)-c1sc3c4c(sc3c1CN(C)C2)-c1sc2c3c(sc2c1CN(C)C4)-c1sc2c4c(sc2c1CN(C)C3)-c1sc(C)cc1CN(C)C4
(31) Fused-S-N-F
136.Cc1cc2c3nc4cc(F)c(F)cc4nc3c3ccsc3c2s1
(32) Fused-S-N-F-O
60.Cc1cc2c(=O)n(C)c(=O)c3c(F)c(C)sc3c2s1
107.Cc1cc2c(=O)n(C)c(=O)c3c(F)csc3c2s1
347.Cc1sc2c(c1F)c(=O)n(C)c(=O)c1c2sc2c1sc1c3scc(F)c3c(=O)n(C)c(=O)c12
(33) Fused-S-N-O
1.Cc1cc2c(=O)n(C)c(=O)c3cc4c(cc3c2s1)c(=O)n(C)c(=O)c1ccsc14
47.Cc1ccc2c(c1)N(C)C(=O)C2=C1C(=O)N(C)c2cc(C)sc21
112.Cc1cc2c(=O)n(C)c(=O)c3c(sc4c3sc3c5sc(C)cc5c(=O)n(C)c(=O)c34)c2s1
121.Cc1ccc2c(c1)N(C)C(=O)C2=C1CCC(C)S1
262.Cc1cc2sc3c(c2s1)N(C)C(=O)C3=C1C(=O)N(C)c2c1sc1cc(C)sc21
274.Cc1cc2c(=O)n(C)c(=O)c3ncsc3c2s1
281.Cc1cc2c3c4c(c5sccc5c5c4c(c2s1)C(=O)N(C)C5=O)C(=O)N(C)C3=O
288.CC1=CC2C(=O)N(C)C(=O)C3N=C(C)SC3C2S1
351.Cc1cc2c(s1)C1=C3C(=O)N(C)C(C)=C3C(=O)N1C=C2
367.CC1=CC2C(=O)N(C)C(=O)C3C=C(C)SC3C2S1
372.CC1=CC2C(=O)N(C)C(=O)C3c4sc5c(c4SC3C2S1)c(=O)n(C)c(=O)c1cc(C)sc15
376.Cc1cc2c(=O)n(C)c(=O)c3nc(C)sc3c2s1
419.Cc1cc2c(=O)n(C)c(=O)c3cc(C)sc3c2s1
422.Cc1nc2c(=O)n(C)c(=O)c3ccsc3c2s1
(34)Fused-S-N-O-Cl
131.CC1C(=O)Nc2c1cc1sccc1c2Cl
222.Cc1cc2c(Cl)c3c(cc2s1)C(C)C(=O)N3C
(35)Fused-S-O
138.Cc1cc2sc(=O)c3ccsc3c2s1
141.Cc1cc2oc(=O)c3ccsc3c2s1
232.Cc1cc2c(s1)-c1sc(C)cc1C(=O)C(C)C2=O
400.Cc1cc2c(s1)-c1cc3c(cc1C2(C)CC(C#C)CO)-c1sc(C)cc1C3

418.Cc1scc2c1C(=O)c1c(C)sc(C)c1C2=O
(36)Fused-S-Si
296.Cc1cc2c(s1)-c1cc3c(cc1[Si]2)-c1sccc1[Si]3
312.Cc1cc2c(s1)-c1sccc1[Si]2
356.Cc1cc2c(s1)-c1sc(C)cc1[Si]2
378.Cc1csc2c1[Si](C)(C)c1sccc1-2
(37)Fused-Si
52.Cc1ccc2c(c1)[Si](C)(C)c1cc(C)ccc1-2
417.CC1=CC2=C(C1)C1=C(C=C(C)C1)[Si]2
(38)Fused-se
182.Cc1cc2c([se]1)-c1[se]c(C)cc1C2
223.Cc1cc2c([se]1)-c1cc3c(cc1C2(C)CCCCCCCCCCCCCCCC)-c1[se]c(C)cc1C3
(39)Mono
17.C#Cc1ccc(C)cc1
54.Cc1ccc(C)c(C)c1
79.CC=Cc1ccc(C)cc1
92.Cc1ccc(C)cc1
126.C=Cc1ccc(C)cc1
160.C#Cc1ccc(C)c(C)c1
188.Cc1ccccc1C
218.CC1=CC=C(C)C1(C)C
220.c1ccccc1
247.Cc1ccccc1
286.C=CC1=CC=C(C)C1
342.CC=Cc1ccc(C)c(C)c1
440.CC#Cc1ccc(C)c(C)c1
442.CC1=C(C)C(C)=C(C)CC1
(40)Mono-F
67.Cc1ccccc1F
72.Cc1cc(F)cc(F)c1
151.Cc1cccc(F)c1F
179.C=Cc1ccc(C)cc1F
194.Cc1c(F)c(F)cc(F)c1F
251.Cc1cccc(F)c1
273.Cc1ccc(C=C)cc1F
303.Cc1c(F)cccc1F
412.Cc1cc(F)ccc1F
443.Cc1ccc(C)c(F)c1
(41)Mono-N
8.Cc1cccnc1
30.Cc1cc(C#N)ccc1C#N
38.Cc1cnc(C)c(C)c1
80.CNc1ccc(C)cc1
86.Cc1ccc(C)nc1
359.CNc1ccc(C)cc1C
360.Cc1cccnn1
383.Cc1ccc(C)cn1
(42)mono-N-O
103.CC1=CC(C)C(=O)N1C

245.CC1=CC(C)C(=O)N1
(43)Mono-O
45.Cc1ccc(C)o1
81.CC1CCC(C)O1
164.Cc1ccc(O)cc1
202.COc1ccc(OC)c(C)c1
254.Cc1cc(C)c(C)o1
302.CC=Cc1ccc(C)c(OC)c1
358.Cc1ccco1
(44)Mono-S
0.CC1CCC(C)S1
32.Cc1cc(CCCC)c(C)s1
36.Cc1csc(C)c1
135.Cc1cc(C)c(C)s1
137.Cc1sccc1C
172.Cc1cccs1
178.Cc1ccsc1C
210.Cc1ccc(C)s1
235.Cc1ccc(C=C)s1
344.C=Cc1ccc(C)s1
353.C#Cc1ccc(C)s1
392.Cc1sc(C=C)cc1C
408.CC=Cc1ccc(C)s1
(45)Mono-S-Cl
259.Cc1cc(Cl)cs1
381.Cc1cc(Cl)c(C)s1
402.Cc1sccc1Cl
(46)Mono-S-F
33.C=Cc1sc(C)cc1F
83.Cc1cc(F)c(C)s1
89.Cc1sc(C)c(C)c1F
187.Cc1cc(F)cs1
228.Cc1scc(F)c1F
391.Cc1sccc1F
394.Cc1sc(C=C)c(F)c1F
(47)Mono-S-N
35.Cc1nc(C)c(C)s1
85.Cc1sccc1C#N
88.Cc1nccs1
132.Cc1nncs1
133.Cc1ncsc1C
170.Cc1cc(C#N)c(C=C)s1
209.Cc1cnc(C)s1
332.CC=NN=Cc1ccc(C)s1
415.Cc1cncs1
421.Cc1sc(C)c(C)c1C#N
(48)Mono-S-N-O
149.C=C1C(=O)N(CC)C(=S)N(CC)C1=O
(49)Mono-S-O

104.CC1CCS(=O)C1
176.CC(=O)Sc1ccc(C)cc1
275.O=S1CCCC1
(50)Mono-se
63.Cc1ccc(C=C)[se]1
161.Cc1ccc(C)[se]1
207.Cc1ccc[se]1
307.C=Cc1ccc(C)[se]1
362.Cc1cc(C)c(C)[se]1

# 2. Polymer-unit types

2.1 Bratch chain combination

'+chain+chain-O+chain-O-Si',
'+chain+chain-N+chain-O+chain-O-Si',
'+chain-F+chain-O+chain-O-Si',
'+chain+chain-O-Si',
'+chain-O-Si',
'+chain+chain-O',
'+chain',
'+chain+chain-N+chain-O',
'+chain-O',
'+chain+chain-N-F+chain-O',
'+chain-O+chain-S-O',
'+chain-O-Si+chain-Si',
'+chain+chain-F',
'+chain-F-O+chain-O',
'+chain+chain-Si',
'+chain-F+chain-O-Si',
'+chain-O+chain-O-Si',
'+chain+chain-O-Si+chain-Si',
'+chain+chain-F+chain-O',
'+chain-F+chain-O',
'+chain+chain-F-O+chain-O',
'+chain-O+chain-S',
'+chain+chain-O+chain-Si',
'+chain+chain-O+chain-S'

2.2 Mono-ring combination

'+mono-se',
'+mono-S',
'+mono-N+mono-S',
'+mono-S+mono-S-F',
'+mono-S-N',
'+mono-S+mono-se',
'no_mono',
'+mono-S-F',
'+mono-F+mono-S',

'+mono-O',
'+mono-S+mono-S-N',
'+mono+mono-se',
'+mono-N+mono-S-F',
'+mono-S-Cl',
'+mono+mono-S',
'+mono',
'+mono-O+mono-S',
'+mono-N-O+mono-S',
'+mono+mono-S-O',
'+mono+mono-N',
'+mono-F',
'+mono-N',
'+mono+mono-O',
'+mono-S+mono-S-O',
'+mono+mono-N+mono-S-N-O',
'+mono-N-O+mono-S+mono-se',
'+mono-S+mono-S-Cl',
'+mono+mono-S-Cl'

## 2.3 Fused/bicyclic ring combination

'+bicyclic-S-N+fused-N-O',
'+bicyclic-N-O',
'+fused+fused-N-O',
'+bicyclic-N-O+fused+fused-S',
'+bicyclic-N-O+fused',
'+fused+fused-S-N-O',
'+fused+fused-N-F-O',
'+bicyclic-S-N-F+fused+fused-N-O',
'+fused+fused-S-N-F-O',
'+bicyclic-N-O+bicyclic-S',
'+fused-N-O',
'+bicyclic-S+fused+fused-N-O',
'+bicyclic-N-O+fused-S-N',
'+bicyclic-S-N+bicyclic-S-N-O',
'+bicyclic-N-O+bicyclic-S-N',
'+bicyclic-N-O-Cl+fused',
'+bicyclic-S+fused+fused-S-N',
'+bicyclic-S-N+fused-S-N-O',
'+bicyclic-N-O+fused+fused-N-O',
'+fused+fused-N-O+fused-S',
'+bicyclic-S-N-F+fused',
'+bicyclic-S-N-F+bicyclic-S-N-O',
'+bicyclic-S-N+fused',
'+fused',
'+bicyclic-S-O+fused',
'+fused+fused-S-N',
'+bicyclic-S-N+bicyclic-S-N-O+fused',
'+bicyclic-N-O+bicyclic-S+fused-N-O',

'+fused-S-N-O',
'+fused+fused-N-O+fused-S-N',
'+fused-S-N-O-Cl',
'+fused+fused-N-O+fused-O',
'+fused+fused-S-O',
'+fused+fused-S-N-O-Cl',
'+bicyclic-S-N+fused+fused-S',
'+fused+fused-S',
'+bicyclic-N-O+bicyclic-S+fused',
'+bicyclic-N-O+bicyclic-S-N+fused',
'+bicyclic+bicyclic-S-N-O+fused',
'+bicyclic-S+bicyclic-S-N-O+fused',
'+bicyclic-S+fused',
'+fused+fused-N',
'+bicyclic-S-N+fused+fused-N',
'+bicyclic-O+bicyclic-S+fused',
'+bicyclic-N-F+fused',
'+bicyclic-S-N-F+fused+fused-S',
'+bicyclic-S-N+fused+fused-S-Si',
'+bicyclic-S-N-F+fused+fused-S-Si',
'+bicyclic-S+bicyclic-S-N-F+fused+fused-S',
'+bicyclic-N+fused',
'+bicyclic-S+bicyclic-S-N+fused+fused-S',
'+bicyclic-S-N+fused+fused-N-O',
'+bicyclic-S-N+fused+fused-se',
'+bicyclic-S-N-F+fused+fused-S-Si+fused-Si',
'+bicyclic-N+bicyclic-N-O+fused',
'+bicyclic-N',
'+bicyclic-N+bicyclic-N-O+bicyclic-S+fused',
'+bicyclic-S+bicyclic-S-O',
'+bicyclic-S-N',
'+bicyclic-N-O+fused+fused-N',
'+bicyclic-S+fused+fused-N-F-O',
'+bicyclic-N-O+bicyclic-S+fused+fused-S-N-O',
'+bicyclic-N-O+fused+fused-S-N-O',
'+bicyclic+bicyclic-N-O+fused',
'+bicyclic-N-O+fused+fused-S-Si',
'no_fused',
'+bicyclic-S-N+fused+fused-O',
'+bicyclic-S+fused+fused-S',
'+bicyclic-N-O+fused-O',
'+bicyclic-N-O+fused-Si',
'+bicyclic-N-O+bicyclic-S-N-O+fused+fused-S-N-O',
'+bicyclic-N+bicyclic-N-O',
'+bicyclic-N-O+bicyclic-S-N-O+fused',
'+bicyclic-S-N+fused+fused-S-O',
'+bicyclic-S-F+fused+fused-S',
'+fused+fused-S+fused-S-O',
'+bicyclic-N-F+fused+fused-S',

'+fused+fused-S+fused-S-N',
'+fused+fused-S+fused-S-N-F',
'+bicyclic-S-N-O+fused',
'+fused+fused-S-Si'

# 3. Parameter Adjustment

MLP_params=[{'hidden_layer_sizes':range(50,150,10),'activation':['identity','logistic','tanh','relu'],'max_iter':range(100,300,10)}]
KNN_params=[{'n_neighbors':range(1,50,1), 'weights':['uniform', 'distance'],'leaf_size':range(1,50,1)}]
gaussian_params=[{'n_restarts_optimizer':range(0,20),'max_iter_predict':range(50,150,10),'multi_class':['one_vs_rest','one_vs_one']}]
SVM_cl_params=[{'C': param_range_svm, 'gamma': np.logspace(-10, 3, 5)}]
RF_params=[{'max_depth': range(1,30,1),'max_features':['auto', 'sqrt', 'log2'], 'n_estimators':range(5,200,5)}]

Complement algorithm:
SGD: SGDClassifier
GDBT: GradientBoostingClassifier
LDA: LinearDiscriminantAnalysis
TREE: DecisionTreeClassifier
SGD_params=[{'loss':['hinge','modified_huber'],'penalty':['l2','l1'],'alpha':[i*0.0001 for i in list(range(1,5,))]}]
GDBT_params=[{'loss':['deviance','exponential'],'n_estimators':range(5,200,5),'criterion':['friedman_mse','squared_error','mse'],'learning_rate':[i*0.1 for i in list(range(1,5))]}]
LDA_params=[{'solver':['svd', 'lsqr', 'eigen'],'store_covariance':['True','False']}]
TREE_params=[{'criterion':['gini','entropy'],'splitter':['best','random'],'max_depth':range(1,30,1)}]

Table.S1 Cross Validation Data-P type

| CV\Algorithm | Gaussian | KNN | MLP | RF | SVM |
|---|---|---|---|---|---|
| 1 | 0.704545455 | 0.75 | 0.704545455 | 0.704545455 | 0.659090909 |
| 2 | 0.659090909 | 0.704545455 | 0.681818182 | 0.681818182 | 0.590909091 |
| 3 | 0.704545455 | 0.727272727 | 0.727272727 | 0.772727273 | 0.704545455 |
| 4 | 0.727272727 | 0.704545455 | 0.772727273 | 0.681818182 | 0.795454545 |
| 5 | 0.681818182 | 0.704545455 | 0.613636364 | 0.681818182 | 0.636363636 |
| 6 | 0.772727273 | 0.681818182 | 0.727272727 | 0.818181818 | 0.75 |
| 7 | 0.681818182 | 0.659090909 | 0.681818182 | 0.727272727 | 0.590909091 |
| 8 | 0.697674419 | 0.674418605 | 0.697674419 | 0.720930233 | 0.697674419 |
| 9 | 0.697674419 | 0.627906977 | 0.674418605 | 0.651162791 | 0.674418605 |
| 10 | 0.744186047 | 0.790697674 | 0.88372093 | 0.790697674 | 0.790697674 |
| mean | 0.707135307 | 0.702484144 | 0.716490486 | 0.723097252 | 0.689006342 |

| CV\Algorithm | SGD | GDBT | LDA | Tree |
|---|---|---|---|---|
| 1 | 0.636363636 | 0.636363636 | 0.568181818 | 0.659090909 |
| 2 | 0.659090909 | 0.75 | 0.659090909 | 0.704545455 |
| 3 | 0.568181818 | 0.704545455 | 0.681818182 | 0.704545455 |
| 4 | 0.613636364 | 0.590909091 | 0.613636364 | 0.613636364 |
| 5 | 0.545454545 | 0.659090909 | 0.590909091 | 0.772727273 |
| 6 | 0.681818182 | 0.727272727 | 0.659090909 | 0.613636364 |
| 7 | 0.681818182 | 0.704545455 | 0.613636364 | 0.704545455 |
| 8 | 0.651162791 | 0.674418605 | 0.604651163 | 0.697674419 |
| 9 | 0.744186047 | 0.790697674 | 0.674418605 | 0.76744186 |
| 10 | 0.720930233 | 0.76744186 | 0.581395349 | 0.744186047 |
| mean | 0.650264271 | 0.700528541 | 0.624682875 | 0.69820296 |

Table.S2 Cross Validation Data-N type

| CV\Algorithm | Gaussian | KNN | MLP | RF | SVM |
|---|---|---|---|---|---|
| 1 | 0.772727273 | 0.727272727 | 0.818181818 | 0.772727273 | 0.681818182 |
| 2 | 0.681818182 | 0.681818182 | 0.772727273 | 0.681818182 | 0.636363636 |
| 3 | 0.681818182 | 0.727272727 | 0.727272727 | 0.681818182 | 0.727272727 |
| 4 | 0.681818182 | 0.681818182 | 0.681818182 | 0.636363636 | 0.636363636 |
| 5 | 0.772727273 | 0.772727273 | 0.772727273 | 0.863636364 | 0.818181818 |
| 6 | 0.590909091 | 0.636363636 | 0.772727273 | 0.590909091 | 0.727272727 |
| 7 | 0.590909091 | 0.636363636 | 0.772727273 | 0.818181818 | 0.772727273 |
| 8 | 0.761904762 | 0.666666667 | 0.761904762 | 0.80952381 | 0.761904762 |
| 9 | 0.80952381 | 0.952380952 | 0.714285714 | 0.714285714 | 0.761904762 |
| 10 | 0.666666667 | 0.761904762 | 0.714285714 | 0.761904762 | 0.714285714 |
| mean | 0.701082251 | 0.724458874 | 0.750865801 | 0.733116883 | 0.723809524 |

| CV\Algorithm | SGD | GDBT | LDA | Tree |
|---|---|---|---|---|
| 1 | 0.636363636 | 0.727272727 | 0.5 | 0.772727273 |
| 2 | 0.545454545 | 0.772727273 | 0.409090909 | 0.681818182 |
| 3 | 0.636363636 | 0.727272727 | 0.590909091 | 0.590909091 |
| 4 | 0.863636364 | 0.863636364 | 0.772727273 | 0.863636364 |
| 5 | 0.636363636 | 0.681818182 | 0.590909091 | 0.818181818 |
| 6 | 0.818181818 | 0.727272727 | 0.545454545 | 0.863636364 |
| 7 | 0.727272727 | 0.636363636 | 0.545454545 | 0.818181818 |
| 8 | 0.545454545 | 0.590909091 | 0.5 | 0.5 |
| 9 | 0.666666667 | 0.666666667 | 0.571428571 | 0.666666667 |
| 10 | 0.714285714 | 0.761904762 | 0.523809524 | 0.714285714 |
| mean | 0.679004329 | 0.715584416 | 0.554978355 | 0.729004329 |

Table.S3 Cross Validation Data-N/P type

| CV\Algorithm | Gaussian | KNN | MLP | RF | SVM |
|---|---|---|---|---|---|
| 1 | 0.781818182 | 0.854545455 | 0.818181818 | 0.763636364 | 0.818181818 |
| 2 | 0.818181818 | 0.8 | 0.872727273 | 0.872727273 | 0.836363636 |
| 3 | 0.8 | 0.763636364 | 0.872727273 | 0.854545455 | 0.872727273 |
| 4 | 0.722222222 | 0.759259259 | 0.87037037 | 0.851851852 | 0.796296296 |
| 5 | 0.777777778 | 0.796296296 | 0.814814815 | 0.833333333 | 0.796296296 |
| 6 | 0.87037037 | 0.907407407 | 0.87037037 | 0.925925926 | 0.888888889 |
| 7 | 0.833333333 | 0.777777778 | 0.87037037 | 0.851851852 | 0.833333333 |
| 8 | 0.851851852 | 0.87037037 | 0.888888889 | 0.851851852 | 0.851851852 |
| 9 | 0.777777778 | 0.740740741 | 0.814814815 | 0.796296296 | 0.796296296 |
| 10 | 0.796296296 | 0.814814815 | 0.833333333 | 0.796296296 | 0.814814815 |
| mean | 0.802962963 | 0.808484848 | 0.852659933 | 0.83983165 | 0.830505051 |

| CV\Algorithm | SGD | GDBT | LDA | Tree |
|---|---|---|---|---|
| 1 | 0.757575758 | 0.878787879 | 0.757575758 | 0.939393939 |
| 2 | 0.666666667 | 0.878787879 | 0.727272727 | 0.848484848 |
| 3 | 0.909090909 | 0.848484848 | 0.666666667 | 0.818181818 |
| 4 | 0.606060606 | 0.909090909 | 0.666666667 | 0.848484848 |
| 5 | 0.909090909 | 0.787878788 | 0.757575758 | 0.757575758 |
| 6 | 0.696969697 | 0.727272727 | 0.787878788 | 0.757575758 |
| 7 | 0.787878788 | 0.787878788 | 0.696969697 | 0.727272727 |
| 8 | 0.787878788 | 0.757575758 | 0.636363636 | 0.757575758 |
| 9 | 0.78125 | 0.71875 | 0.625 | 0.75 |
| 10 | 0.75 | 0.875 | 0.78125 | 0.875 |
| mean | 0.765246212 | 0.816950758 | 0.713352273 | 0.807954545 |

# 4. data supplement

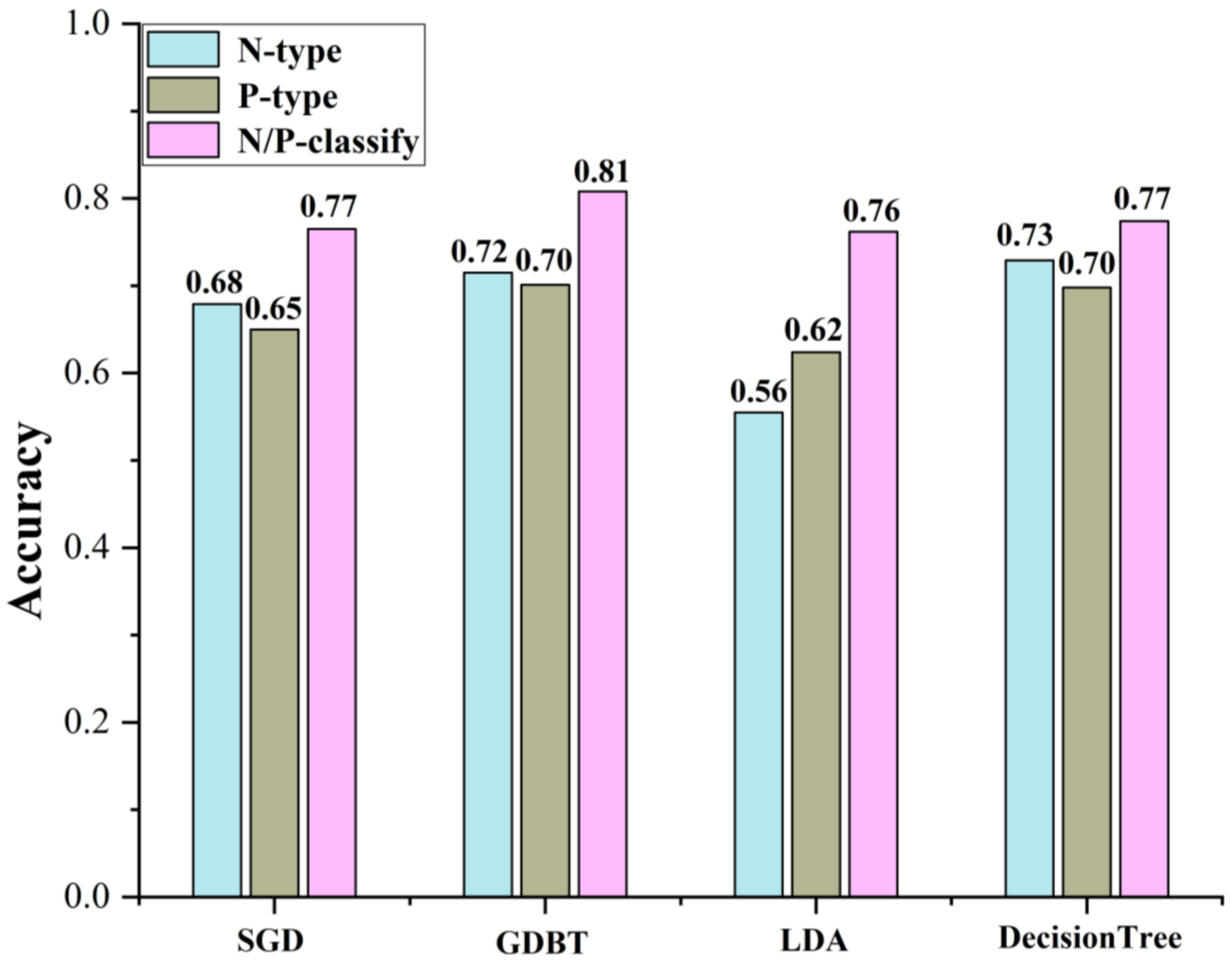

Fig.S1 Other machine learning methods

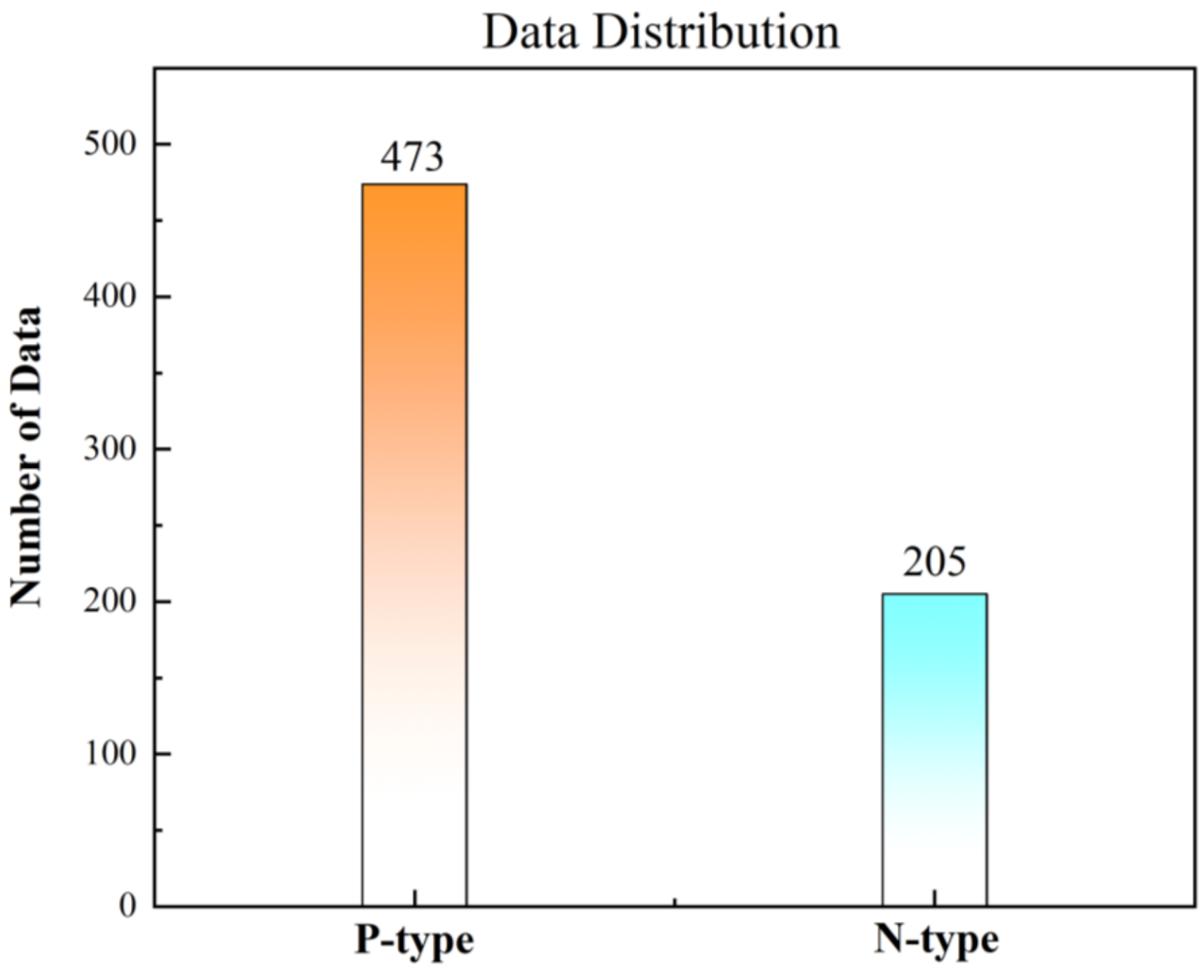

Fig. S2 Distribution of data

Given that the data distribution does affect the prediction results, we randomly sampled 205 of the 473 P-type data to make the amount of N-type and P-type data equal. Machine learning prediction of N/P classification was made in this dataset. The accuracy slipped slightly, but the maximum was still above 80%(Fig.S3).

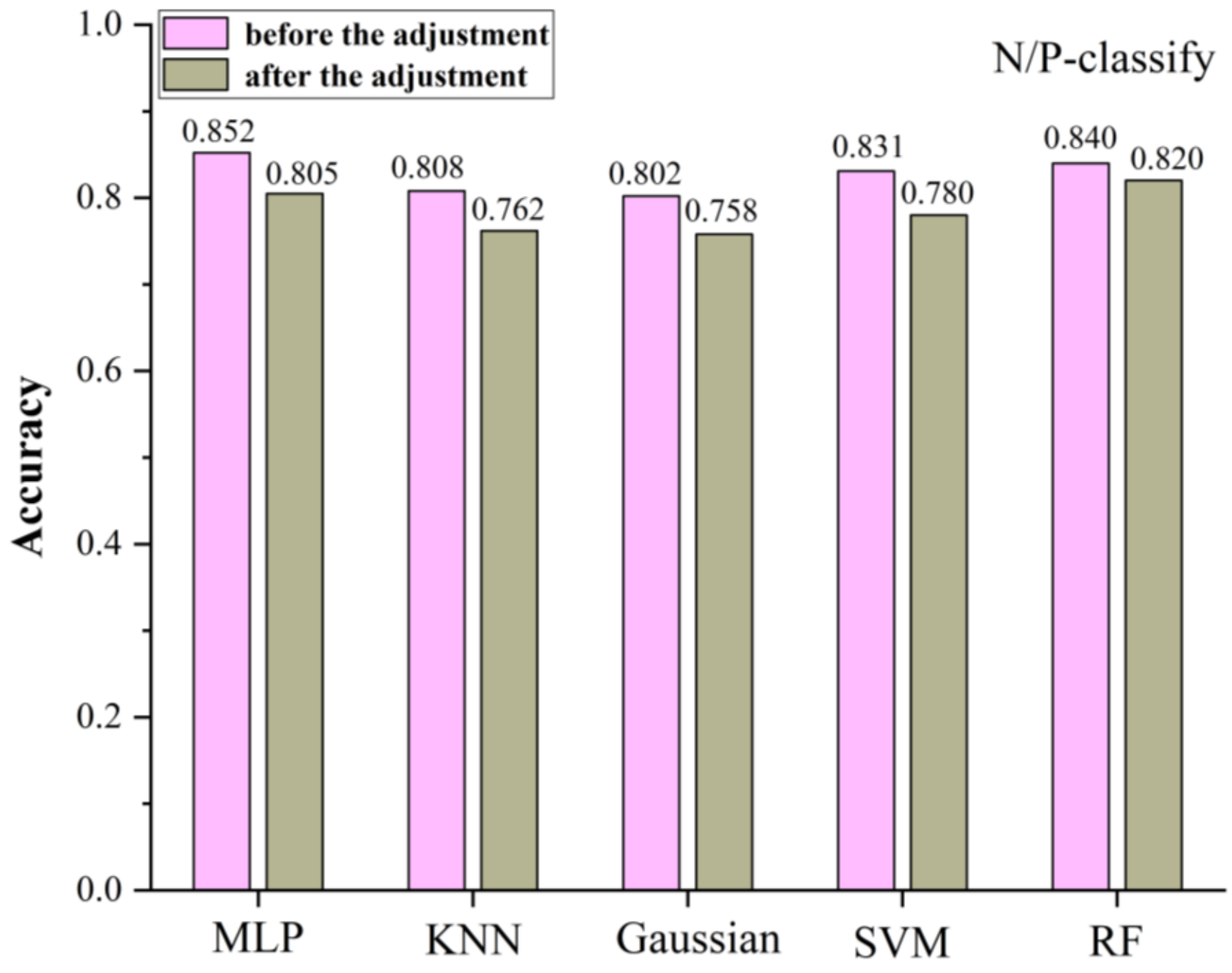

Fig. S3 N/P classify after data adjustment

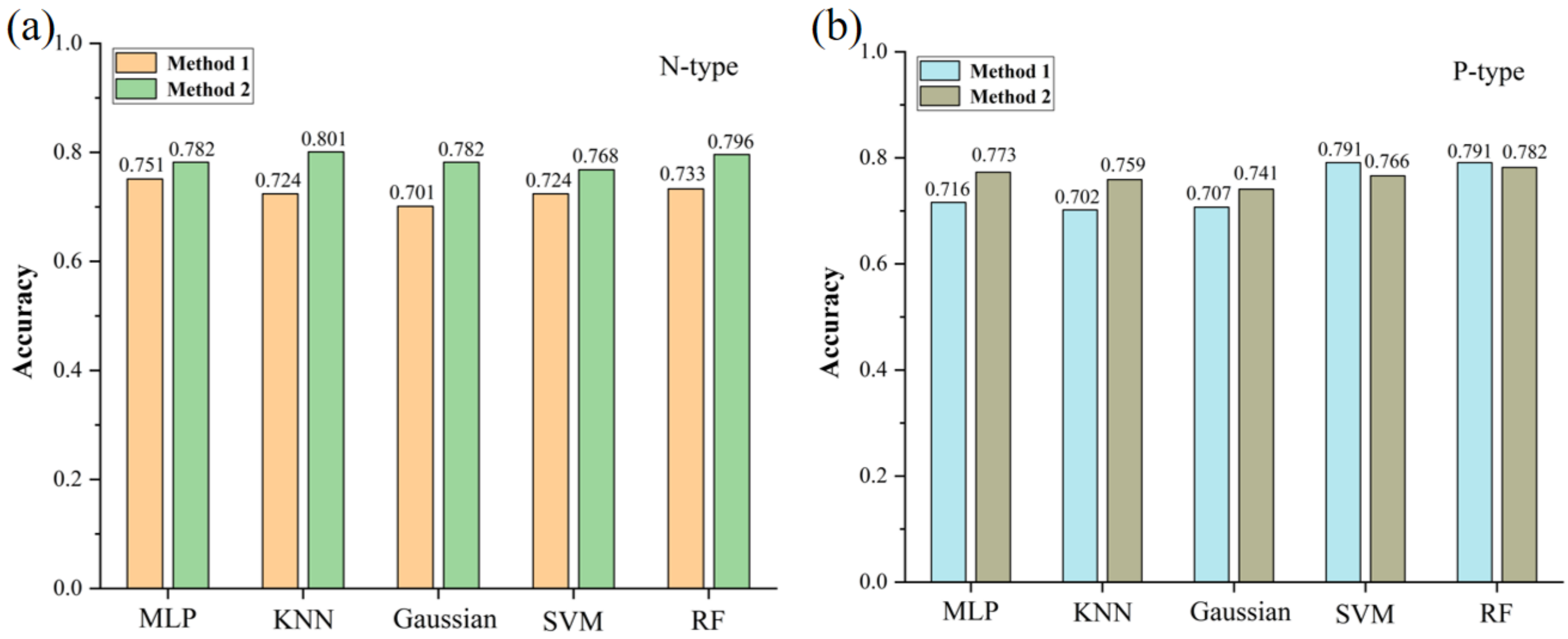

Fig.S4 Change the classification method:(a) N-type,(b)P-type.

Method1: The classification method used in the manuscript-- The mobility is classified as high (> 4 cm2·V-1·s-1), medium (1 to 4 cm2·V-1·s-1), or low (0 to 1 cm2·V-1·s-1).

Method2:

$$\mu \rightarrow \mathrm{int}(\ln(\mu+1)) \qquad \text{e.q.S1}$$

The carrier mobility is transformed by e.q.S1.